\documentclass[12pt]{article}

\usepackage{amssymb}
\usepackage{color}
\usepackage{epsfig,amssymb,amsfonts,amsmath,graphicx,dsfont,cite,xfrac}
\usepackage{authblk}
\usepackage{subfigure}

\definecolor{mygray}{gray}{0.5}

\usepackage{cite}
\usepackage[colorlinks=true,linkcolor=blue,citecolor=red]{hyperref}

\newcommand{\be}{\begin{equation}}
\newcommand{\ee}{\end{equation}}
\newcommand{\bea}{\begin{eqnarray}}
\newcommand{\eea}{\end{eqnarray}}

\title{Transition from the wave equation to either the heat or the transport equations through fractional differential expressions}

\author[${}$]{Fernando Olivar-Romero}
\author[${}$]{Oscar Rosas-Ortiz}

\affil[${}$]{\footnotesize Physics Department, Cinvestav, AP 14-740, 07000
M\'exico City, Mexico}

\date{}
\begin{document}

\maketitle

\begin{abstract}

We study a model that intermediates among the wave, heat, and transport equations. The approach considers the propagation of initial disturbances in  a one-dimensional medium that can vibrate. The medium is nonlinear in such a form that nonlocal differential expressions are required to describe the time-evolution of solutions. Non-locality is modeled with a space-time fractional differential equation of order $1 \leq \alpha \leq 2$ in time, and order $1\leq \beta \leq 2$ in space. We adopt the notion of Caputo for time-derivative and the Riesz pseudo-differential operator for space-derivative. The corresponding Cauchy problem is solved for zero initial velocity and initial disturbance represented by either the Dirac delta or the Gaussian distributions. Well known results for the partial differential equations of wave propagation, diffusion and (modified) transport processes are recovered as particular cases. In addition, regular solutions are found for the partial differential equation that arises from $\alpha=2$ and $\beta=1$. Unlike the above cases, the latter equation permits the presence of nodes in its solutions.

\end{abstract}


\section{Introduction}

Partial differential equations play a relevant role in mathematical physics, mainly those of the second order \cite{Tik63,Gus99,Duf15,Bor18}. In analogy to conics of analytic geometry, they are classified in either elliptic, parabolic or hyperbolic. These equations are useful to describe a diversity of physical phenomena that include wavelike propagation, diffusion and transport processes in practically all the branches of physics. Namely, in continuous and classical mechanics it is common to face problems of either the hyperbolic (vibrating strings, stretched membranes) or the parabolic (heat conduction) types. In statistical mechanics, to describe the motion of a cloud of noninteracting particles, it is necessary to solve a transport (linear Boltzmann) equation, which is of parabolic type, and so on. However, as a fingerprint of the above models, the local character of the differential operators used to construct the corresponding equations permeates over the predictions. The latter produces solutions with unpleasant features from time to time because, occasionally, the predicted behavior is not compatible with the observed phenomenon. For example, the heat equation predicts that thermal disturbances propagate at infinite velocity. Clearly, the model offered by such equation is unrealistic, although it works well when diffusivity is slower than the velocity of propagation. Other examples include the study of systems that exhibit the combination of two (or more) traits. Viscoelasticity, for instance, is understood as the combination of viscous and elastic behavior exhibited by   some materials when they are deformed. In such cases the local profile of conventional differential operators represents a limitation to model the involved phenomena in realistic fashion.

A form to remove the intricacies of local operators consists in substituting the usual concept of linear medium (where the phenomenon under study is observed) by the idea of nonlinear medium with memory. This approach has been successfully applied to change the unpleasant infinite propagation velocity of conventional diffusion by a more realistic finite velocity in e.g. \cite{Gur68,Mil78}. The main point is that memory effects may be modeled with nonlocal differential equations that arise from the action of fractional differential operators on the appropriate set of functions. The same approach is useful to investigate viscoelasticity, where the fractional calculus finds a great deal of applications \cite{Mai10}.

This work is addressed to study a model that intermediates among the wave, heat, and transport equations. That is, we are interested in connecting pure wavelike propagation with pure diffusion  and  transport processes by means of the fractional differential properties of a single model in unified form. Our approach considers the propagation of initial disturbances in a one-dimensional medium that can vibrate. When the medium is disturbed at a given time by a vibration, perturbations spread out from the disturbance at a given velocity. The medium is considered nonlinear in such a form that non-locality is required in both, space and time variables. Thus, we are going to study a space-time fractional differential equation with time-derivative of order $1 \leq \alpha \leq 2$, and space-derivative of order $1 \leq \beta \leq 2$. For the time-derivative we shall adopt the one introduced by Caputo \cite{Cap67}, the space-derivative will be defined by the Riesz pseudo-differential operator \cite{Rie49}. The values taken by $\alpha$ and $\beta$ define a squared area in the $\alpha \times \beta$-plane that will parameterize the solutions of our model. Interestingly, we find regular solutions for the differential equation that arises by fixing $\alpha=2$ and $\beta=1$. Unlike the conventional equations aforementioned, this new equation permits the presence of nodes in its solutions. The zeros appear in pairs and propagate, together with the maxima and minima of the solutions, in wavelike form. As far as we know, such equation as well as its solutions have been unexplored in the literature up to now. In this form, the zoology of combined phenomena may be enlarged by involving the above equation and its solutions.

The organization of the paper is as follows. Motivated by the non-locality of fractional differential equations, in Section~\ref{section2} we formulate a Cauchy problem that embraces, as particular cases,  three well known problems of mathematical physics: to solve either the (hyperbolic) wave equation, the (parabolic) heat equation or the (modified) transport equation for properly defined initial conditions. The differential equation associated to $\alpha=2$ and $\beta=1$ is a modified version of an equation of the  parabolic type which completes the set of second order partial differential equations that can be constructed by omitting the term of simultaneous space and time derivatives. Section~\ref{section3} includes the solution for two different Cauchy problems, both of them considering zero initial velocity. The first one is defined by the Dirac delta distribution as the initial disturbance. The solutions are written in terms of the $H$-function, their  convergence is analyzed and concrete expressions are given for particular cases that include the well known results of conventional differential equations. Some directrices about the form of intertwining such equations are also given. The second problem is defined by considering a Gaussian distribution as the initial disturbance. We show that the solution is a convergent series of $H$-functions. To our knowledge, this result has been missing in the literature up to now. We also show that such a series goes to the solution of the Dirac delta case at the appropriate limit. In Section~\ref{results} we include the analysis of our results and discuss about the form in which the maxima of the solutions propagate in the $t \times x$-plane. In accordance with other studies, we find that the fractional order of the differential equations affects the behavior of the characteristic integrals in such a form that they are not straight-lines anymore. However, our results show that the time-dependence of the maxima is not as simple as this has been reported by other authors. Some concluding remarks are included in Section~\ref{conclu}. To offer fluidity in reading the manuscript, we have resigned detailed calculations of important solutions to appendices~\ref{ApB}, \ref{ApC}, and \ref{ApD}. In turn, Appendix~\ref{ApA} includes useful information and expressions that are recursively used throughout the paper.

\section{Problem formulation} 
\label{section2}

The one-dimensional wave equation \cite{Tik63,Gus99,Duf15,Bor18}
\be
\left[ \frac{\partial^{2} }{\partial t^{2}} - v^{2}\frac{\partial^{2} }{\partial x^{2}}\right] u(x,t)=0,
\label{we} 
\ee
with the initial conditions 
\be
u(x,0)=\varphi_{1}(x),  \quad  \left. \frac{\partial u(x,t)}{\partial t} \right\vert_{t=0} =\varphi_{2}(x),
\label{ini1}
\ee
admits a unique solution given by the D'Alembert's formula 
\be
u(x,t)= \frac{\varphi_1(x+vt) + \varphi_1(x-vt)}{2} + \frac{1}{2v} \int_{x-vt}^{x+vt} \varphi_2(z) dz.
\label{dal}
\ee
The Cauchy's problem (\ref{we})-(\ref{ini1}) is defined for $t > 0$, with sufficiently smooth functions $\varphi_1$ and $\varphi_2$. The first term on the right-hand side of (\ref{dal}) represents the propagation of the initial wavelike disturbance $\varphi_1$, ruled by the {\em characteristic} integrals $x \pm vt = \rm{const}$, for zero initial velocity $\varphi_2=0$. Hence, the positive parameter  $v$ represents the velocity of propagation of the wave. Consistently, $\varphi_1(x-vt)$ propagates to the right and $\varphi_1(x+vt)$ to the left. In turn, the integral term of (\ref{dal}) corresponds to vibrations produced by the initial velocity $\varphi_2$ with no initial disturbance $\varphi_1=0$. 

A case of special interest is defined by the Dirac delta distribution \cite{Duf15} as follows
\be
\varphi_1^{(\delta)} (x) = \mu \delta (x), \quad \varphi_2(x)=0.
\label{ini2}
\ee
The initial disturbance (\ref{ini2})  is a very localized pulse that levels off rapidly and has a strength equal to $\mu$. Therefore
\be
u_{\delta}(x,t)= \frac{\mu}{2} \left[ \delta(x+vt) + \delta(x-vt) \right]
\label{sol1}
\ee
is solution of the one-dimensional wave equation (\ref{we}) with the initial conditions (\ref{ini2}).

On the other hand,  to define uniquely the solution of the one-dimensional heat equation \cite{Tik63,Gus99,Duf15,Bor18,Wid75}
\be
\left[ \frac{\partial}{\partial t} - k \frac{\partial^{2} }{\partial x^{2}}\right] u(x,t)=0,
\label{he} 
\ee
it is generally sufficient the initial condition \cite{Wid75}
\be
u(x,0)= \varphi_1(x),
\label{ini3}
\ee
where $u(x,t)$ is the absolute temperature and $k>0$ stands for the {\em thermometric conductivity} (or {\em diffusivity}) \cite{Wid75}. Any line parallel to the $x$-axis is a {\em characteristic} integral of Eq.~(\ref{he}). The Cauchy problem (\ref{he})-(\ref{ini3}) is useful to describe diffusion processes without drift in one-dimension. If the initial function (\ref{ini3}) is the Dirac delta distribution then (\ref{he}) is the Fokker-Planck (or forward Kolmogorov) equation \cite{Uma15}. Indeed, it may be shown that the time-dependent Gaussian density
\be
u_S(x,t)= 
\frac{\mu}{\sqrt{4\pi kt}}e^{-x^{2}/4kt}, \quad t > 0,
\label{source}
\ee
often called {\em source solution} \cite{Tik63}, satisfies (\ref{he}) with the initial condition 
\be
\varphi_1^{(\delta)} (x) = \mu \delta(x).
\label{ini4}
\ee 
Clearly, density (\ref{source}) converges to the initial distribution (\ref{ini4}) at the limit $t \rightarrow 0^+$. Explicitly
\be
\lim_{t \rightarrow 0^+} u_S(x,t) = \infty, \quad \lim_{t \rightarrow 0^+} \int_{-\epsilon}^{\epsilon} u_S(x,t) dx = \mu, \quad \epsilon >0. 
\label{delta}
\ee
Besides, at fixed time, $u_S(x,t)$ falls off very rapidly as $\vert x \vert$ increases. Therefore, density (\ref{source}) is useful to describe diffusion in one-dimensional media where the heat is concentrated in the vicinity of $x=0$ \cite{Bor18,Wid75}. In addition, $u_S(x,t)$ is positive and nonzero for any $x$ and $t $, so one also finds that  ``the effect of introducing a quantity of heat at $x=0$ is instantaneously noticeable at remote points'' \cite{Wid75}. That is, thermal disturbances propagate at infinite velocity. Although unrealistic, this model works well in situations where diffusivity is slower than the speed of propagation. If diffusivity and the speed of thermal waves are compatible then (\ref{he}) must be modified to include an additional second order time-derivative term \cite{Duf15}.

One of the directions of this work is addressed to study a model that intermediates between the wave  and the heat equations described above. That is, we look for a connection between pure wave propagation and pure diffusive phenomena in one-dimension. A first resource consists of replacing the second order time-derivative of the wave equation (\ref{we}) by a fractional time-derivative $D^{\alpha}$ of order $\alpha \in [1,2]$. In this case we have
\be
\left[ D^{\alpha} - v^2_{\alpha} \frac{\partial^2 }{\partial x^2 } \right] u(x,t) =0, \quad 1 \leq \alpha \leq 2.
\label{fracwe} 
\ee
Here, $v_{\alpha} >0$ is a constant expressed in the units $[v_{\alpha}] =[L] [T]^{-\alpha/2}$, with $[L]$ and $[T]$ standing for length and time units. Using $\alpha=2$ in (\ref{fracwe}) we recover the wave equation (\ref{we}) while $\alpha=1$ gives the heat equation (\ref{he}), with $v_{\alpha=2} =v$ and $v_{\alpha = 1}^2=k$ respectively. Remarkably, the unpleasant feature of the heat equation (\ref{he}) that thermal disturbances propagate at infinite velocity is removed since (\ref{fracwe}) describes heat conduction for nonlinear materials with memory, where the speed of propagation is finite \cite{Gur68,Mil78}. Extending the values of $\alpha$ to include the sub-diffusion interval $0 < \alpha < 1$, one arrives at a fractional version of the Fokker-Planck equation for the initial condition (\ref{ini4}). However, such a case is out of the scope of the present work and will be studied elsewhere. Preliminary results associated to (\ref{fracwe}) have been already reported by other authors in e.g.  \cite{Wys86,Sch89,Fuj90,Mai96b,Mai96,Mai01} (see also the general approaches discussed in \cite{Mai10,Uma15,Gor14,Pod99,Her18,Uch13}). 

To include a major diversity of cases we may replace the second order space-derivative of the heat equation (\ref{he}) by a fractional space-derivative $\mathbb{D}^{\beta}$ of order $\beta \in [1,2]$ to obtain
\be
\left[ \frac{\partial}{\partial t} - k_{\beta}  \mathbb{D}^{\beta} \right] u(x,t)=0, \quad 1 \leq \beta \leq 2.
\label{frache} 
\ee
The constant $k_{\beta} >0$, measured in units $[k_{\beta}]= [L]^{\beta} [T]^{-1}$, is such that $k_2=k$. Equation (\ref{frache}) is useful to describe symmetric L\'evy-Feller diffusion processes in which jump components are present \cite{Uma15,Gor99} (see also \cite{Sai97,Gor98}, and general approaches in \cite{Mai10,Uma15,Gor14,Pod99,Her18,Uch13}). The parameter $\beta=2$ gives the heat equation and $\beta=1$ leads to a modified version of the one-dimensional transport equation \cite{Gor00,Luc13}
\be
\left[ \frac{\partial}{\partial t} - k_1 \mathbb D^1 \right] u(x,t)=0.
\label{te}
\ee
It may be shown that for any smooth function $f(x)$, the expression $u(x,t) =f(x + k_1 t)$ solves the conventional transport equation; therefore, for the initial condition (\ref{ini3}) we have $u(x,t) = \varphi_1(x + k_1 t)$, where $k_1$ is the {\em fluid} ({\em hydrodynamic}) velocity \cite{Med10}. The latter is not automatic for Eq.~ (\ref{te}), see \cite{Gor00,Luc13}. For the sake of simplicity, Eq.~(\ref{te}) will be referred to as transport equation.

\begin{figure}[htb]
\centering 
\includegraphics[width=0.3\textwidth]{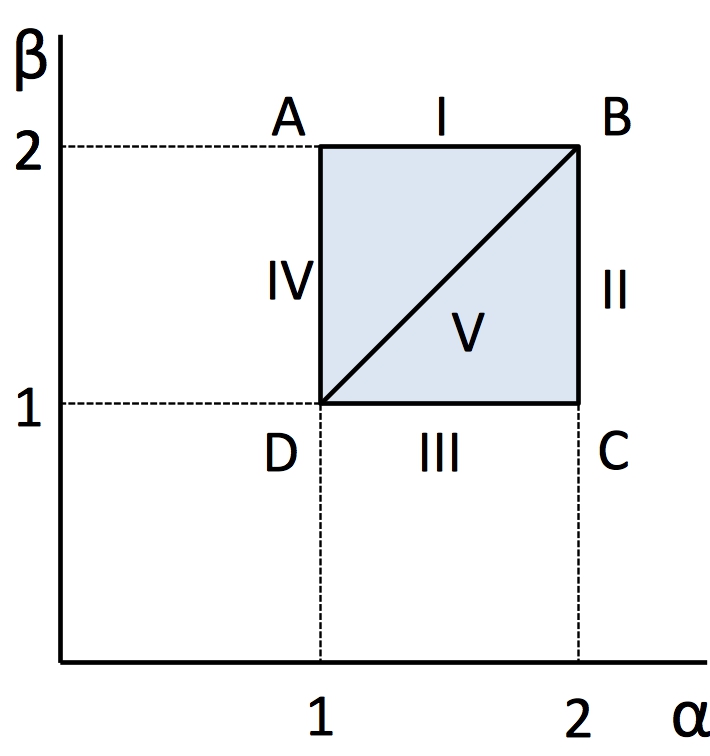} 

\caption{\footnotesize The $\alpha \times \beta$-plane associated to the space-time differential equation (\ref{frac1}). The square $1 \leq \alpha, \beta \leq 2$ contains the points $(\alpha, \beta)$ that are the subject of interest in this work. Vertices $B$, $A$ and $D$ correspond to the wave equation (\ref{we}), the heat equation (\ref{he}), and the transport equation (\ref{te}), respectively. The solutions of (\ref{frac1}) for different initial conditions (\ref{ini1}) associated to the grey squared area are defined within the text. Special attention is paid to the five segment-lines defined by $\alpha = 1,2$, $\beta = 1,2$, $\alpha=\beta$, and the vertices $A,B,C,D$. See complementary information in Table~\ref{table1}.
}
\label{plane}
\end{figure}

The most general alternative corresponds to substitute both derivatives, the temporal and the spatial ones, by their fractional versions in Eq.~(\ref{we}). Thus, we have the space-time fractional differential equation \cite{Gor00,Luc13}
\be
\left[ D^{\alpha}  -v_{\alpha,\beta}^{2} \mathbb{D}^{\beta} \right] u(x,t) = 0, \quad 1 \leq \alpha \leq 2, \quad 1 \leq \beta \leq 2,
\label{frac1} 
\ee
where the coupling constant $v_{\alpha,\beta} >0$ is written in the units $[v_{\alpha,\beta}] = [L]^{\beta/2}[T]^{-\alpha/2}$, and is such that $v_{\alpha,2} \equiv v_{\alpha}$, $v_{1, \beta}^2 = k_{\beta}$, and $v_{2,1}^2 = \kappa$. The above equation can be studied in different approaches \cite{Mai10,Uma15,Gor14,Pod99,Uch13,Her18}, and offers very interesting mathematical features as the fact that one can transit from the wave equation (\ref{we}) to the heat one (\ref{he}), and vice versa, by fixing $\beta =2$ and running $\alpha$ in $[1,2]$, see segment $BA$ in Figure~\ref{plane}. In turn, the heat equation (\ref{he}) serves as point of departure to arrive at the transport equation (\ref{te}) by fixing $\alpha =1$ and permitting $\beta$ to take values in $[1,2]$, see segment $AD$ in Figure~\ref{plane}. The transition from the transport equation (\ref{te}) to the wave equation (\ref{we}) is reached by making $\alpha=\beta$, with $\alpha \in [1,2]$, see segment $DB$ in Figure~\ref{plane}. Of course, any other (differentiable) path described by the point $(\alpha, \beta)$ to connect $B$ with either $A$ or $D$ is admissible in the $\alpha \times \beta$-plane. As indicated above, we are interested in those paths described within the grey squared area of Figure~\ref{plane}, and the segment-lines $I$-$V$ are representative of the simplest ones. In this respect, the zone enclosed by the circuit $BAD$ refers to phenomena that intermediate among wave propagation, diffusive and transport processes. The zone delimited by the circuit $BCD$ represents an additional resource of information. We shall discuss the subject in the next sections.

In addition to the differential equations (\ref{we}) and (\ref{he}), together with Eq.~(\ref{te}), we remark the differential form
\be
\left[ \frac{\partial^2 }{\partial t^2} - \kappa \mathbb D^1\right] u(x,t)=0,
\label{Ceq}
\ee
which is defined by the space-time fractional differential equation (\ref{frac1}) evaluated at vertex $C$, and will be called {\em complementary equation}. This can be achieved from either the wave or the transporttransport equations, (\ref{we}) and (\ref{te}), through segment-lines $II$ and $III$, respectively. In this form, the circuit $ABCD$ is completed as the combination of triangles $BAD$ and $BCD$ of Figure~\ref{plane}. 

As we can see, the diversity of points $(\alpha,\beta)$ that can be used to define a concrete form of Eq.~(\ref{frac1}) embraces important cases of the family of second-order partial differential equations
\be
F(x,t,u,u_x, u_t, u_{xx}, u_{tt})=0,
\label{family}
\ee
where the sub-labels of $u$ refer to partial derivatives in conventional notation. The term ``complementary'' coined for Eq.~(\ref{Ceq}) is then justified. In the language of linear second order partial differential equations the set (\ref{family}) includes only hyperbolic and parabolic types. Indeed, the wave equation (\ref{we}) is hyperbolic while the heat equation (\ref{he}) is parabolic. We shall say that complementary and transport equations, (\ref{Ceq}) and (\ref{te}) respectively, are also ``parabolic''.

For the time-derivative $D^{\alpha}$ we shall adopt the one introduced by Caputo \cite{Cap67}. In turn, the Riesz pseudo-differential operator $\mathbb{D}^{\beta}$ \cite{Rie49} is going to substitute the space-derivatives in our approach.

\section{Solution and examples} 
\label{section3}

In this section we study the Cauchy problem defined by the space-time fractional differential equation (\ref{frac1}) and the initial conditions (\ref{ini1}). In general, we assume that the solutions $u(x,t ; \alpha, \beta): \mathbb{R} \times (0,\infty) \rightarrow \mathbb{R}$ will represent the propagation of initial wavelike disturbances $\varphi_1$ for the initial velocity $\varphi_2$ in a medium that can vibrate. That is, when the medium is disturbed at a given (initial) time by a vibration  $\varphi_1$, perturbations spread out from the disturbance at initial velocity $\varphi_2$. From now on we omit the explicit dependence of $u$ on the parameters $\alpha$ and $\beta$ for the sake of simplicity. It is expected that $u$ will converge either to the solutions of the wave equation (\ref{we}) or to the solutions of the heat equation (\ref{he}) at the limits $(\alpha, \beta) \rightarrow (2,2)$ and $(\alpha, \beta) \rightarrow (1,2)$, respectively. The same holds at the limit $(\alpha, \beta) \rightarrow (1,1)$ for the transport equation (\ref{te}). Our main interest is addressed to the values $1 < \alpha, \beta <2$, with either $\alpha=\beta$ or $\alpha \neq \beta$. In Figure~\ref{plane} we show the $\alpha \times \beta$-plane associated to the functions $u(x,t)$ we are interested in.

Let us start by calculating the Laplace transform $\mathfrak{L}$ of Eq.~(\ref{frac1}) in the time-variable. After introducing the initial conditions (\ref{ini1}) we arrive at the new equation
\be
s^{\alpha}U(x,s)-s^{\alpha-1}\varphi_{1}(x)-s^{\alpha-2}\varphi_{2}(x)=v_{\alpha,\beta}^{2} \mathbb D^{\beta}U(x,s), 
\label{frac2} 
\ee
where $U(x,s)$ stands for the Laplace transform of $u(x,t)$. Now, we calculate the Fourier transform $\mathfrak{F}$ of Eq.~(\ref{frac2}) in the space-variable. We obtain 
\be
s^{\alpha} \overline{U} \left( k,s \right) - s^{\alpha-1\,} \overline\varphi_{1}(k) - s^{\alpha-1\, } \overline\varphi_{2}(k) = - v_{\alpha,\beta}^{2} \vert k \vert^{\beta \, }\overline{U}\left(k,s\right),
\label{frac3}
\ee
which is immediately solved by the function
\be
\overline{U} \left(k,s\right) = \frac{s^{\alpha-1 \,} \overline\varphi_{1}(k) + s^{\alpha-2 \, } \overline\varphi_{2}(k) } {s^{\alpha}+ v_{\alpha,\beta}^{2} \vert k \vert^{\beta} }.
\label{frac4}
\ee
In the above equations we have introduced the notation $\mathfrak{F}[\varphi_{i}(x)]=\overline\varphi_{i}(k)$, $i=1,2$, and $\overline{U} \left(k,s\right) = \mathfrak{F} [U(x,s) ]$. To retrieve $U$ we calculate  the inverse Fourier transform
\be
U(x,s) = \frac{1}{\sqrt{2\pi} } \int_{\mathbb R} e^{ikx} \, \overline U(k,s) dk.
\label{lap1}
\ee
Providing the initial disturbance $\varphi_1$ and velocity $\varphi_2$, the solution we are looking for is obtained from the inverse Laplace transform $\mathfrak{L}^{-1}$ of (\ref{lap1}). That is, $u(x,t) = \mathfrak{L}^{-1} [ U(x,s)]$. Next, we determine the explicit form of $u(x,t)$ for $\varphi_2=0$ and  two different functions $\varphi_1$.

\subsection{Delta-like disturbances} 
\label{iniciales1}

For initial conditions (\ref{ini2}), the integral (\ref{lap1}) is simplified as follows
\be
U (x,s) = \frac{\mu}{2\pi} \int_{\mathbb R} \,  \frac{s^{\alpha-1} e^{ikx} }{s^{\alpha}+v_{\alpha,\beta}^{2} \vert k \vert^{\beta}} \,\, dk.
\label{lap2}
\ee
The straightforward calculation (see details in Appendix~\ref{ApB}) gives
\be 
U(x,s)= \left(  \frac{ \mu \, s^{\alpha/\beta-1} }{ \beta \, v_{\alpha,\beta}^{2/\beta}} \right) 
H_{2,3}^{2,1} \left[ \left. \left( \frac{s^{\alpha/\beta}}{v_{\alpha,\beta}^{2/\beta}} \right) \vert x \vert \,\right\rvert 
\begin{array}{cc} 
\left( \frac{\beta -1}{\beta},\frac{1}{\beta} \right), \left( \frac{1}{2},\frac{1}{2} \right) \\[1.5ex]
(0,1), \left( \frac{\beta -1}{\beta},\frac{1}{\beta} \right), \left( \frac{1}{2},\frac{1}{2} \right) 
\end{array}
\right], 
\label{udes} 
\ee
where $H_{p,q}^{m,n}[z|-] \equiv H(z)$ stands for the Fox function, $H$-function for short \cite{Kil04,Mat10} (see also Appendix~\ref{ApA}). The absolute value of $x$ in (\ref{udes}) obeys parity properties of the Fourier transform. The full derivation of the inverse Laplace transform of (\ref{udes}) is included in Appendix~\ref{ApC}, we obtain
\be
u(x,t)= \left( \frac{\mu}{\beta \, t^{\frac{\alpha}{\beta} } \, v_{\alpha,\beta}^{2/\beta} } \right) H_{3,3}^{2,1} \left[ \left. \frac{ \vert x \vert }{t^{\frac{\alpha}{\beta} } \, v_{\alpha,\beta}^{2/\beta} } \,  \right\rvert 
\begin{array}{cc} 
\left( \frac{\beta -1}{\beta}, \frac{1}{\beta} \right),  \left( \frac{1}{2}, \frac{1}{2} \right), \left( \frac{ \beta - \alpha}{\beta}, \frac{\alpha}{\beta} \right) \\[1.5ex]
(0,1), \left( \frac{\beta - 1}{\beta},\frac{1}{\beta} \right), \left( \frac{1}{2}, \frac{1}{2} \right) 
\end{array} 
\right]. 
\label{udelta} 
\ee
Equation (\ref{udelta}) represents the propagation of the disturbance produced by a very localized pulse $\varphi_1^{ (\delta) }(x)$ in one-dimensional media, according to the space-time fractional differential equation (\ref{frac1}).

\subsubsection{Discussion about the convergence of solutions} 
\label{discussion}

To analyze the convergence of the functions $u(x,t)$ defined in (\ref{udelta}) one may consider the series expansion of the $H$-function given in Theorems~1.3 and 1.4 of Ref.~\cite{Kil04} (see Appendix~\ref{ApA}). It can be distinguished two cases. For $\alpha > \beta$ (points delimited by the triangle $BCD$ of Figure~\ref{plane}) we have the series
\be
u(x,t)= \frac{\mu}{\sqrt{\pi} |x|} \sum_{k=1}^{\infty} \frac{ \Gamma ( \frac{1+ \beta k}{2} )}{
\Gamma (1+ \alpha k) \Gamma (- \frac{\beta}{2} k) }
\left( -\frac{ 2^{\beta}  v_{\alpha,\beta}^2 \, t^{\alpha} }{\vert x \vert^{\beta}} \right)^k, \quad x \neq 0,
\label{serie1}
\ee
which converges absolutely \cite{Gor00}. On the other hand, for $\beta>\alpha$ (points bordered by the triangle $BAD$ of Figure~\ref{plane}), the function $u(x,t)$ acquires the form
\bea
u(x,t) &=& \frac{ \mu \vert x \vert^{\beta-1} }{ \sqrt{\pi} 2^{\beta} t^{\alpha} v_{\alpha,\beta}^{2} } 
\sum_{k=0}^{\infty} \frac{ \Gamma ( \frac12 - \frac{\beta}{2} -\frac{\beta }{2} k) }  
{ \Gamma (1-\alpha - \alpha k) \Gamma ( \frac{\beta}{2}+\frac{\beta }{2} k) }
\left( -\frac{ |x|^{\beta} } {2^{\beta}t^{\alpha}v_{\alpha,\beta}^{2} } \right)^k 
\nonumber\\[1.5ex]
&& + \frac{\mu}{\sqrt{\pi} \beta t^{\frac{\alpha}{\beta} } v_{\alpha,\beta}^{2/\beta} } \sum_{m=0}^{\infty}  
\frac{ \Gamma (\frac{1+2m}{\beta} ) \Gamma ( 1 -\frac{1+2m}{\beta} )}{ m! \, 
\Gamma (\frac12 +m) \Gamma \left( 1 - \frac{\alpha (1+2m) }{\beta} \right) } \left( 
-\frac{x^2}{ 4 t^{\frac{2\alpha}{\beta} } v_{\alpha,\beta}^{4/\beta} } \right)^m.
\label{serie2}
\eea
Thus, any point $(\alpha, \beta)$ in the zone defined by the circuit $BCD$ produces a convergent solution (\ref{udelta}). In turn, for the points $(\alpha, \beta)$ within the triangle $BAD$, the convergence of $u(x,t)$ must be determined from the series (\ref{serie2}). For instance, as $\alpha=1$ produces the cancelation of the first term on the right-hand side of (\ref{serie2}), we arrive at the following expression for the points on segment-line $IV$ of Figure~\ref{plane}, 
\be
u(x,t) = \frac{\mu}{\pi \beta t^{\frac{1}{\beta} } k_{\beta}^{1/\beta} } \sum_{m=0}^{\infty}  
\frac{ \Gamma (\frac{1+2m}{\beta} ) }{ \Gamma (2m)  } \left( 
- \frac{x^2}{ t^{\frac{2}{\beta} } k_{\beta}^{2/\beta} } \right)^m, \qquad   1< \beta \leq 2, \quad \alpha=1.
\label{serie3}
\ee
Remark that,  in agreement with the results reported in \cite{Gor00}, Eq.~(\ref{serie3}) do not include the 
point $(\alpha, \beta) = (1,1)$ associated to vertex $D$.

The case $\alpha = \beta$ (segment-line $V$ of Figure~\ref{plane}) intermediates between the two series aforementioned. The straightforward calculation \cite{Gor00} shows that $u(x,t)$ is reduced to the expression
\be
u(x,t) = \left( \frac{\mu}{\pi} \right)  \frac{ \vert x \vert^{\alpha-1} v_{\alpha,\alpha}^2  t^{\alpha} \sin (\pi \alpha/2) }{ 
v_{\alpha,\alpha}^4 t^{2\alpha} + 2 \vert x \vert^{\alpha} v_{\alpha, \alpha}^2 t^{\alpha} \cos (\pi \alpha/2) + \vert x \vert^{2\alpha}  }, \quad \alpha=\beta.
\label{serie4}
\ee

\subsubsection{Special cases} 
\label{special}

We find the following special cases of function (\ref{udelta}). The sub-label of $u$ refers to the diagram shown in Figure~\ref{plane}. For solutions along the segment-lines $I$-$V$ we have
\bea
u_I(x,t) &=& \left( \frac{\mu}{2 \, t^{\frac{\alpha}{2} } \, v_{\alpha,2} } \right) H_{1,1}^{1,0} \left[ \left. \frac{ \vert x \vert }{t^{\frac{\alpha}{2} } \, v_{\alpha,2} } \,  \right\rvert 
\begin{array}{cc} 
\left( \frac{ 2 - \alpha}{2}, \frac{\alpha}{2} \right) \\[1.5ex]
(0,1)
\end{array} 
\right]
\nonumber
\\[1.5ex]
& = & \left( \frac{\mu}{2 \, t^{\frac{\alpha}{2} } \, v_{\alpha,2} } \right) \phi \left( -\frac{\alpha}{2}, 1 - \frac{\alpha}{2};  - \frac{ \vert x \vert }{t^{\frac{\alpha}{2} } \, v_{\alpha,2} }
\right),
\label{u1}
\eea
where $\phi(a,b;z)$ is the Wright function (see Appendix~\ref{ApA}). The above expression is a well known result for Eq.~(\ref{fracwe}) with initial conditions (\ref{ini2}), see e.g. \cite{Gor00}. We also have
\be
u_{II} (x,t)= \left( \frac{\mu}{\beta \, t^{\frac{2}{\beta} } \, v_{2,\beta}^{2/\beta} } \right) H_{3,3}^{2,1} \left[ \left. \frac{ \vert x \vert }{t^{\frac{2}{\beta} } \, v_{2,\beta}^{2/\beta} } \,  \right\rvert 
\begin{array}{cc} 
\left( \frac{\beta -1}{\beta}, \frac{1}{\beta} \right),  \left( \frac{1}{2}, \frac{1}{2} \right), \left( \frac{ \beta - 2}{\beta}, \frac{2}{\beta} \right) \\[1.5ex]
(0,1), \left( \frac{\beta - 1}{\beta},\frac{1}{\beta} \right), \left( \frac{1}{2}, \frac{1}{2} \right) 
\end{array} 
\right],
\label{u2} 
\ee

\be
u_{III}(x,t)= \left( \frac{\mu}{ t^{\alpha}  v_{\alpha,1}^2} \right) H_{3,3}^{2,1} \left[ \left. \frac{ \vert x \vert }{t^{\alpha} v_{\alpha,1}^2 } \,  \right\rvert 
\begin{array}{cc} 
(0,1),  \left( \frac{1}{2}, \frac{1}{2} \right), ( 1 - \alpha, \alpha)\\[1.5ex]
(0,1), (0,1), \left( \frac{1}{2}, \frac{1}{2} \right) 
\end{array} 
\right].
\label{u3} 
\ee
According to Section~\ref{discussion}, the above solutions are absolutely convergent series. However, although $u_{II}$ and $u_{III}$ give rise to $u_B$ and $u_D$ respectively, they are scarcely explored in the literature. On the other hand, both of them converge to the solution associated to vertex $C$. That is, they lead to the solution of the complementary equation (\ref{Ceq}) which, in turn, seems to be missing in the set of exactly solvable differential equations that are commonly studied in physics and enginery (see for instance the monograph \cite{Uch13}).

The function
\be
u_{IV}(x,t)= \left( \frac{\mu}{\beta \, t^{\frac{1}{\beta} } \, k_{\beta}^{1/\beta} } \right) H_{2,2}^{1,1} \left[ \left. \frac{ \vert x \vert }{t^{\frac{1}{\beta} } \, k_{\beta}^{1/\beta} } \,  \right\rvert 
\begin{array}{cc} 
\left( \frac{\beta -1}{\beta}, \frac{1}{\beta} \right),  \left( \frac{1}{2}, \frac{1}{2} \right) \\[1.5ex]
(0,1), \left( \frac{1}{2}, \frac{1}{2} \right) 
\end{array} 
\right]
\label{u4} 
\ee
converges to $u_A(x,t)$ and $u_D(x,t)$ as $\beta \rightarrow 2$ and $\beta \rightarrow 1$, respectively. As solution of the space-differential equation (\ref{frache}), the expression (\ref{u4}) is in correspondence with results that have been already reported in e.g. \cite{Uma15,Gor99,Sai97,Gor98}. The comparison of (\ref{u4}) with (\ref{serie3}) gives rise to the following evaluation of the $H$-function
\[
H_{2,2}^{1,1} \left[  \vert z \vert \left\rvert 
\begin{array}{cc} 
\left( \frac{\beta -1}{\beta}, \frac{1}{\beta} \right),  \left( \frac{1}{2}, \frac{1}{2} \right) \\[1.5ex]
(0,1), \left( \frac{1}{2}, \frac{1}{2} \right) 
\end{array} 
\right.
\right] = \frac{1}{\pi} \sum_{m=0}^{\infty}  (-1)^m
\frac{ \Gamma (\frac{1+2m}{\beta} ) }{ \Gamma (2m)  }  z^{2m}, \quad 1 < \beta \leq 2.
\]

The function
\be
u_V(x,t)= \left( \frac{\mu}{\beta  t   v_{\beta,\beta}^{2/\beta} } \right) H_{2,2}^{1,1} \left[ \left. \frac{ \vert x \vert }{t  v_{\beta,\beta}^{2/\beta} } \,  \right\rvert 
\begin{array}{cc} 
\left( \frac{\beta -1}{\beta}, \frac{1}{\beta} \right),  \left( \frac{1}{2}, \frac{1}{2} \right) \\[1.5ex]
\left( \frac{\beta - 1}{\beta},\frac{1}{\beta} \right), \left( \frac{1}{2}, \frac{1}{2} \right) 
\end{array} 
\right]
\label{u5} 
\ee
embraces the results for $\alpha=\beta$ reported in e.g. \cite{Luc13,Gor00}, and regulates the simplest transition from $u_D(x,t)$ to $u_B(x,t)$ and vice versa. Comparing (\ref{u5}) with (\ref{serie4}) we obtain another evaluation of the $H$-function. Namely
\be
H_{2,2}^{1,1} \left[  \frac{ \vert x \vert }{\gamma} \,  \left\rvert 
\begin{array}{cc} 
\left( \frac{\beta -1}{\beta}, \frac{1}{\beta} \right),  \left( \frac{1}{2}, \frac{1}{2} \right) \\[1.5ex]
\left( \frac{\beta - 1}{\beta},\frac{1}{\beta} \right), \left( \frac{1}{2}, \frac{1}{2} \right) 
\end{array} 
\right.
\right] = \left( \frac{\gamma \beta}{\pi} \right) \frac{ \vert z \vert^{\beta-1} \gamma^{\beta} \sin(\pi \beta/2)}{ 
\gamma^{2\beta} + 2 \vert z \vert^{\beta} \gamma^{\beta} \cos(\pi \beta/2) + \vert z \vert^{\beta} }, \quad \gamma >0.
\ee

The solutions defined by the four vertices can be written as
\be
u_A(x,t) = \left( \frac{\mu}{2  \sqrt{kt} } \right) H_{1,1}^{1,0} \left[ \left. \frac{ \vert x \vert }{ \sqrt{kt} } \,  \right\rvert 
\begin{array}{cc} 
\left( \frac12, \frac12 \right) \\[1.5ex]
(0,1)
\end{array} 
\right] 
 = \left( \frac{\mu}{2  \sqrt{kt} } \right) \phi \left( -\frac12, \frac12; -\frac{ \vert x \vert }{ \sqrt{kt} } \right).
\label{ua}
\ee
Compare Eq.~(\ref{ua}) with $u_I(x,t)$ and $u_{IV}(x,t)$. The function
\bea
u_B(x,t) &=& \left( \frac{\mu}{2  vt } \right) H_{1,1}^{1,0} \left[ \left. \frac{ \vert x \vert }{ vt} \,  \right\rvert 
\begin{array}{cc} 
(0,1)\\[1.5ex]
(0,1)
\end{array} 
\right]  \equiv
\left( \frac{\mu}{2  vt } \right) H_{2,2}^{1,1} \left[ \left. \frac{ \vert x \vert }{ vt} \,  \right\rvert 
\begin{array}{cc} 
\left( \frac12, \frac12 \right)\\[1.5ex]
\left( \frac12, \frac12 \right)
\end{array} 
\right]
\nonumber
\\[1.5ex]
&=& \left( \frac{\mu}{2  vt } \right) \phi \left(-1,0; - \frac{ \vert x \vert }{ vt} 
\right)
\label{ub}
\eea
can be recovered from either $u_I(x,t)$, $u_{II}(x,t)$ or $u_V(x,t)$. For the functions
\be
u_C(x,t)= \left( \frac{\mu}{ \kappa t^2} \right) H_{3,3}^{2,1} \left[ \left. \frac{ \vert x \vert }{ \kappa t^2 } \,  \right\rvert 
\begin{array}{cc} 
(0,1),  \left( \frac{1}{2}, \frac{1}{2} \right), ( -1, 2)\\[1.5ex]
(0,1), (0,1), \left( \frac{1}{2}, \frac{1}{2} \right) 
\end{array} 
\right],
\label{uc} 
\ee
and
\be
u_D(x,t)= \left( \frac{\mu}{  k_1 t} \right) H_{2,2}^{1,1} \left[ \left. \frac{ \vert x \vert }{ k_1 t } \,  \right\rvert 
\begin{array}{cc} 
(0,1), \left( \frac{1}{2}, \frac{1}{2} \right) \\[1.5ex]
(0,1), \left( \frac{1}{2}, \frac{1}{2} \right) 
\end{array} 
\right],
\label{ud} 
\ee
see $u_{II}$, $u_{III}$, and $u_{III}$, $u_{IV}$ and $u_V$, respectively. In Table~\ref{table1} we summarize the above results.

\begin{table}[htb]
\begin{center}
\begin{tabular}{ccc}
\hline
Fractional Expression  & Simplest Transition & Involved Equations\\
\hline
$u_I(x,t)$ & $u_B \leftrightarrow u_A$ & wave $\leftrightarrow$ heat\\[.5ex]
Eq. (\ref{u1}) & Eq. (\ref{ub}) $\leftrightarrow$ Eq. (\ref{ua}) & Eq. (\ref{we}) $\leftrightarrow$ Eq. (\ref{he}) \\
\hline
$u_{IV}(x,t)$ & $u_A \leftrightarrow u_D$ & heat $\leftrightarrow$ transport\\[.5ex]
Eq. (\ref{u4}) & Eq. (\ref{ua}) $\leftrightarrow$ Eq. (\ref{ud}) & Eq. (\ref{he}) $\leftrightarrow$ Eq. (\ref{te}) \\
\hline
$u_V (x,t)$ & $u_D \leftrightarrow u_B$ & transport $\leftrightarrow$ wave\\[.5ex]
Eq. (\ref{u5}) & Eq. (\ref{ud}) $\leftrightarrow$ Eq. (\ref{ub}) & Eq. (\ref{te}) $\leftrightarrow$ Eq. (\ref{we}) \\
\hline
$u_{II}(x,t)$ & $u_B \leftrightarrow u_C$ & wave $\leftrightarrow$ compl.\\[.5ex]
Eq. (\ref{u2}) & Eq. (\ref{ub}) $\leftrightarrow$ Eq. (\ref{uc}) & Eq. (\ref{we}) $\leftrightarrow$ Eq. (\ref{Ceq}) \\
\hline
$u_{III}(x,t)$ & $u_C \leftrightarrow u_D$ & compl. $\leftrightarrow$ transport\\[.5ex]
Eq. (\ref{u3}) & Eq. (\ref{uc}) $\leftrightarrow$ Eq. (\ref{ud}) & Eq. (\ref{Ceq}) $\leftrightarrow$ Eq. (\ref{te}) \\
\hline
\end{tabular}
\end{center}
\caption{\footnotesize The four differential equations defined by Eq.~(\ref{frac1}) for the vertices $A,B,C,D$ can be intertwined by paths described in the squared area $1 \leq \alpha, \beta \leq 2$. The simplest transition from one of them to any other is achieved through the solutions (\ref{udelta}) associated to the five segment lines $I$-$V$ shown in Figure~\ref{plane}.
}
\label{table1}
\end{table}


\subsubsection{Recovering conventional results} 

{\bf Wave equation.} To evaluate $u(x,t)$ at the limit $(\alpha,\beta)\to(2,2)$ we use the solution (\ref{ub}) associated to  vertex~B. Of course, $u_I$, $u_{II}$, and $u_V$ are also useful at the appropriate limits. Using Eqs.~(\ref{H}) and (\ref{H2}) of Appendix~\ref{ApA} we get
\be
u_B (x,t) = \left(  \frac{\mu}2  \right) \frac1{2\pi i} \int_L (vt)^{z-1} \vert x \vert^{-z} dz.
\label{lim1}
\ee
From (\ref{a3}) of Appendix~\ref{ApA} we see that the above integral corresponds to the inverse Mellin transform of $\mu \delta (\vert x \vert -vt)/2$. Then
\be
u_B (x,t) = \left(  \frac{\mu}2  \right)  \delta (\vert x \vert -vt).
\label{lim2}
\ee
This last expression reproduces the solution $u_{\delta}(x,t)$ given in (\ref{sol1}) for the one dimensional wave equation (\ref{we}) with initial conditions (\ref{ini2}).

$\bullet$ As collateral result, the combination of (\ref{ub}) and (\ref{lim2}) gives rise to the following evaluation of the $H$-function:
\be
H_{1,1}^{1,0} \left[ \left. \frac{\vert x \vert }{\gamma}  \, \right\rvert 
\begin{array}{cc} 
(0,1)\\[1.5ex]
(0,1)
\end{array} 
\right] = \phi \left(-1,0; - \frac{\vert x \vert}{\gamma} \right)=
\gamma \delta( \vert x \vert - \gamma), \quad \gamma >0.
\label{hdelta}
\ee

\noindent
{\bf Heat equation.} In similar form, the limit $(\alpha,\beta)\to(1,2)$ is obtained from the solution (\ref{ua}) associated to vertex~A. Equivalent results are obtained from either $u_I$ or $u_{IV}$ at the appropriate limits. We have
\be
u_A(x,t) = \left( \frac{\mu}{2\sqrt{kt}} \right) \frac{1}{2\pi i} \int_{L}\frac{ \Gamma \left( z \right) }{ \Gamma \left( \frac{1+z}{2} \right) } \left( \frac{ \vert x \vert }{\sqrt{kt}} \right)^{-z}dz.
\ee
Using the duplication formula of the Gamma function (\ref{dupli}), as well as the Cahen-Mellin integral (\ref{CMf}), we arrive at the function $u_S(x,t)$ defined in (\ref{source}), which is solution of the one-dimensional heat equation (\ref{he}) with initial condition (\ref{ini4}). The verification that $u_A(x,t)$ satisfies the initial condition (\ref{ini4}) is easily achieved by using Eq.~(\ref{delta}).

$\bullet$ In this case, the combination of (\ref{ua}) and (\ref{source}) gives rise to the evaluation
\be
H_{1,1}^{1,0} \left[ \left. \frac{\vert x \vert}{\gamma } \, \right\rvert 
\begin{array}{cc} 
\left( \frac12, \frac12 \right) \\[1.5ex]
(0,1)
\end{array} 
\right] = \phi \left( -\frac12, \frac12; -\frac{ \vert x \vert}{\gamma} \right)= 
\frac{e^{-x^2/4\gamma^2} }{\sqrt \pi}, \quad \gamma >0.
\label{hgauss1}
\ee
From (\ref{delta}) we also have
\be
\lim_{\gamma \rightarrow 0^+}  \frac{1}{ 2 \gamma }
H_{1,1}^{1,0} \left[ \left. \frac{\vert z \vert}{\gamma^{1/2}} \, \right\rvert 
\begin{array}{cc} 
\left( \frac12, \frac12 \right) \\[1.5ex]
(0,1)
\end{array} 
\right] = \lim_{\gamma \rightarrow 0^+}  \frac{1}{ 2 \gamma } \phi \left( -\frac12, \frac12; -\frac{ \vert x \vert}{\gamma} \right)=
\delta(x).
\label{hgauss2}
\ee

\noindent
{\bf Transport equation.} The limit $(\alpha,\beta)\to(1,1)$ is evaluated from (\ref{ud}), associated to vertex~D (other options include either $u_{III}$ or $u_{IV}$ at the appropriate limit). That is, from (\ref{H}) and (\ref{H2}) of Appendix~\ref{ApA}, after using the duplication formula (\ref{dupli}), we have
\be
u_D(x,t)= \left( \frac{\mu}{ \pi k_1 t} \right) \frac1{2 \pi i} \int_L 
\frac12 \, \Gamma \left( \frac{z}{2} \right) \Gamma \left( 1 - \frac{z}{2} \right) 
\left( \frac{\vert x \vert }{  k_1 t} \right)^{-z} dz.
\label{antes}
\ee
From the Mellin transformation (\ref{b31}) of Appendix~\ref{ApB}, one gets
\[
{\cal M} [ \tfrac{1}{1+y^2}]= \tfrac12 \, \Gamma \left( \tfrac{z}{2} \right) \Gamma \left( 1 - \tfrac{z}{2} \right).
\]
Then, (\ref{antes}) is reduced to the curve
\be
u_D (x,t) = \frac{\mu}{ \pi }  \left[ \frac{ k_1 t }{x^2 + (k_1 t)^2}
\right],
\label{fock}
\ee
which is consistent with Eq.~(\ref{serie4}). The latter result is well known in the literature (see, e.g. \cite{Gor00}, Remark~3.2). In the space variable, $u_D(x,t)$ describes a bell-shaped curve known as either Cauchy (mathematics), Lorentz (statistical physics) or Fock-Breit-Wigner (nuclear and particle physics) distribution \cite{Ros08}. It is centered at $x=0$ (the {\em location parameter}), with a half-width at half-maximum equal to $k_1t$ (the {\em scale parameter}) and amplitude (height) equal to $\mu (\pi k_1 t)^{-1}$. That is, at time $t$, the fluid velocity $k_1$ defines the width of the disturbance between the half-maximum points $x =\pm k_1t$. The function $u_D(x,t)$ satisfies the Dirac delta constraints (\ref{delta}), which can be verified at the elementary level. Therefore, (\ref{fock}) satisfies the transport equation (\ref{te}) with initial condition (\ref{ini4}).

$\bullet$ The combination of (\ref{ud}) with (\ref{fock}) gives the evaluation
\be
H_{2,2}^{1,1} \left[ \left. \frac{ \vert z \vert }{ \gamma} \,  \right\rvert 
\begin{array}{cc} 
(0,1), \left( \frac{1}{2}, \frac{1}{2} \right) \\[1.5ex]
(0,1), \left( \frac{1}{2}, \frac{1}{2} \right) 
\end{array} 
\right] = \frac{1}{\pi} \left( \frac{ \gamma }{z^2 + \gamma^2} \right), \quad \gamma >0,
\ee
with the following limit
\be
\lim_{\gamma \rightarrow 0^+} \gamma^{-1} H_{2,2}^{1,1} \left[ \left. \frac{ \vert z \vert }{ \gamma} \,  \right\rvert 
\begin{array}{cc} 
(0,1), \left( \frac{1}{2}, \frac{1}{2} \right) \\[1.5ex]
(0,1), \left( \frac{1}{2}, \frac{1}{2} \right) 
\end{array} 
\right] =  \delta (z ), \quad \gamma >0.
\ee

\noindent
{\bf Complementary equation.} For $\alpha=2$ and $\beta=1$ the series (\ref{serie1}) is simplified as follows
\be
u_C(x,t) = -\frac{2\mu \kappa}{\sqrt{\pi} } \left( \frac{t}{\vert x \vert} \right)^2
\sum_{n=0}^{\infty} \frac{ \Gamma(n+1)}{ \Gamma(4n+3) \Gamma(-n-\frac12) } \left( \frac{2 \kappa t^2}{ \vert x \vert } \right)^{2n}, \quad x \neq 0.
\label{csol}
\ee
Depending on the value of $ \xi =t^2/\vert x \vert >0$, the alternation of sign in $\Gamma(-n-\frac12)$, $n=0,1,\ldots$, produces pairs of zeros in $u_C(x,t)$ that are located symmetrically with respect to $x=0$. As an immediate example, for $\xi <<1$ the series (\ref{csol}) coincides with the peaked density 
\be
u_C(x,t) \approx \frac{\mu \kappa }{2 \pi } \left( \frac{t}{\vert x \vert} \right)^2
, \quad x\neq 0,
\label{peak}
\ee
which is consistent with the initial condition $\varphi_1^{(\delta)}(x)$. The number of zeros increases as $\xi \rightarrow \infty$. We will find a first pair at second order of $\xi^2$, and so on. On the other hand, $u_C(x,t)$ goes to zero from above as $\vert x \vert \rightarrow \infty$, see Eqs.~(\ref{csol})-(\ref{peak}). So that the zeros produced during the propagation of the pulse $\varphi_1^{(\delta)} (x)$ are nodes of $u_c(x,t)$.

\subsubsection{Other representations}

It is well known that the $H$-function embraces a large number of functions of the hypergeometric type \cite{Mat10}. Among the functions preferred in fractional calculus, besides the $H$-function, one finds Wright and Mittag--Leffler functions \cite{Mai10,Pod99,Her18,Gor14}. To show the flexibility of our results, in this section we rewrite the function (\ref{udelta}) in terms of the generalized Wright function. The translation to other hypergeometric representations is straightforward. 

After changing $z$ by $1-s$, and using the duplication formula of the Gamma function (\ref{dupli}) in the Mellin-Barnes integral representation (\ref{c1}) of $u(x,t)$, we have 
\be
u(x,t)= \frac{\mu}{\sqrt{\pi}\beta |x|} \frac{1}{2\pi i} 
\int_L\frac{ 
\Gamma ( \tfrac{s}{\beta} ) 
\Gamma ( 1- \frac{s}{\beta} ) 
\Gamma ( \tfrac12 - \frac{s}{2} ) 
}{ \Gamma ( 1-\tfrac{\alpha }{\beta} s ) \Gamma ( \frac{s}{2} ) }
\left( \frac{ 2  v_{\alpha,\beta}^{2/\beta} \, t^{\frac{\alpha}{\beta}}}{|x|} \right)^{-s} ds.
\label{25}
\ee
The change of variable $s=\beta y$ produces
\be
u(x,t)= \frac{\mu}{\sqrt{\pi}  |x|} \frac{1}{2\pi i} 
\int_L\frac{ 
\Gamma ( y ) \Gamma ( 1- y ) \Gamma ( \tfrac12 - \frac{\beta}{2} y ) 
}{ \Gamma ( 1-\alpha y ) \Gamma ( \frac{\beta}{2} y) }
\left( \frac{ 2^{\beta}  v_{\alpha,\beta}^2 \, t^{\alpha} }{ \vert x \vert^{\beta} } \right)^{-y} dy.
\label{disc1}
\ee
Using Eqs.~(\ref{H2})-(\ref{H}) of Appendix~\ref{ApA} one can write
\be
u(x,t)= \frac{\mu}{\sqrt{\pi} |x|} H_{2,3}^{1,2} \left[ 
\left.  \frac{ 2^{\beta}  v_{\alpha,\beta}^2 \, t^{\alpha} }{ \vert x \vert^{\beta} } \,  \right\rvert 
\begin{array}{cc} 
( 0, 1), ( \frac12, \frac{\beta}{2} ) \\[1.5ex]
( 0, 1 ), ( 0, \alpha ), ( 1, -\frac{\beta}{2} )
\end{array} 
\right].
\label{disc2}
\ee
Now, from Eq.~(\ref{HPsi}) of Appendix~\ref{ApA} we finally arrive at the function
\be
u (x,t)= \frac{\mu}{\sqrt{\pi} |x| } \,   {}_2 \Psi_2 \left[ 
 -\frac{ 2^{\beta}  v_{\alpha,\beta}^2 \, t^{\alpha} }{\vert x \vert^{\beta} }\,  \left\rvert 
\begin{array}{cc} 
( 1, 1 ), ( \frac{1}{2}, \frac{\beta}{2} ) \\[1.5ex]
( 1, \alpha ), ( 0, -\frac{\beta}{2} )
\end{array} 
\right. \right].
\label{disc3}
\ee
The Mellin-Barnes representation of $u(x,t)$ given in (\ref{25}) coincides with the expression of the Green function reported in \cite{Gor00} for the Cauchy problem defined by Eq.~(\ref{frac1}) and the initial conditions $u(x,0)= \varphi_1$, $\varphi_2=0$. In agreement with our results, in \cite{Gor00} it is found that the concrete form of $u(x,t)$ given in (\ref{disc3}) is reduced to the absolutely convergent series (\ref{serie1}) for $\alpha > \beta$. Other values of $\alpha$ and $\beta$ lead to the expressions discussed in Section~\ref{discussion} and Ref.~\cite{Gor00}.


\subsection{Gaussian disturbances} 
\label{iniciales2}

The Gaussian density (\ref{source}) offers a very versatile profile to define the initial condition $\varphi_1$. Namely, at arbitrary time $t=t_0>0$, we can take $x_0= 2 \sqrt{kt_0}$ to write
\be
\varphi_1^{ (G)} (x)= \mu \frac{e^{-(x/x_0)^2} }{ x_0 \sqrt{\pi} }.
\label{inig}
\ee
What makes density (\ref{inig}) interesting as initial condition is that, unlike the Dirac delta distribution $\varphi_1^{(\delta)} (x)$, this is finite at $x=0$ for $x_0 \neq 0$. The configuration studied in the previous sections for $\varphi_1^{(\delta)} (x)$ can be recovered from (\ref{inig}) at the limit $x_0 \rightarrow 0^+$, as it is clear from Eq.~(\ref{delta}). Additionally, we have shown that the Gaussian density (\ref{inig}), as well as the Dirac delta pulse (\ref{ini4}), can be expressed in terms of the $H$-function, Eqs.~(\ref{hgauss1}) and (\ref{hgauss2}) respectively. The full derivation of the solution of Eq.~(\ref{frac1}) with initial conditions (\ref{inig}) and $\varphi_2=0$ is given in Appendix~\ref{ApD}. The final result can be written as the series
\be
u_e(x,t)= \frac{\mu}{ \beta t^{\frac{\alpha}{\beta} } \, v_{\alpha,\beta}^{2/\beta} } \sum_{k=0}^{\infty} \frac{(-1)^k }{k!} \left( \frac{x_0^2}{ 4 t^{\frac{2\alpha}{\beta} } \, v_{\alpha,\beta}^{4/\beta} } \right)^k  \Theta_k(z;\alpha,\beta),
\label{solg1}
\ee
where
\be
\Theta_k(z;\alpha,\beta) = H_{3,3}^{2,1} \left[ \left. \frac{ \vert x \vert }{ 
t^{\frac{\alpha}{\beta} } \, v_{\alpha,\beta}^{2/\beta}  } \right\rvert 
\begin{array}{cc} 
\left( \frac{\beta -(1 + 2k)}{\beta}, \frac{1}{\beta} \right),
(\frac12, \frac12),
\left( \frac{\beta-\alpha (1+2k) }{\beta}, \frac{\alpha}{\beta} \right)\\[1.5ex]
(0,1), \left( \frac{\beta -(1+2k) }{\beta}, \frac{1}{\beta} \right), 
(\frac12, \frac12) 
\end{array}
\right].
\label{solg2}
\ee
Remarkably, with exception of the term with $k=0$, the coefficients of the series (\ref{solg1})-(\ref{solg2}) become zero at the limit $x_0 \rightarrow 0^+$. Therefore we arrive at the expression
\be
\lim_{x_0 \rightarrow 0^+} u_e (x,t) = \left( \frac{\mu}{ \beta t^{\frac{\alpha}{\beta} } \, v_{\alpha,\beta}^{2/\beta} } \right) \Theta_0 (z;\alpha,\beta),
\ee
which coincides with the solution (\ref{udelta}) of the space-time fractional differential equation (\ref{frac1}) with initial conditions (\ref{ini2}), as expected.

\section{Analysis of the results}
\label{results}

\begin{figure}[htb]

\centering
\subfigure[ $(1.1, 1.95 )$ ]{\includegraphics[width=0.26\textwidth]{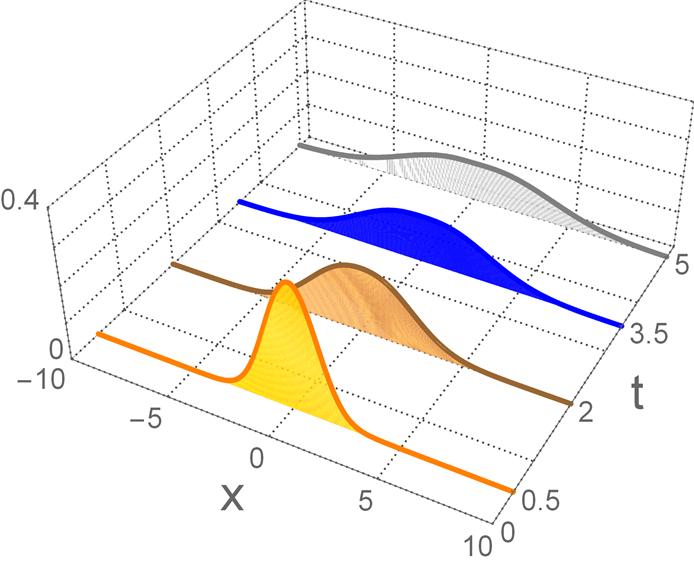} } 
\hskip.5ex
\subfigure[  $(1.5, 1.95 )$ ]{\includegraphics[width=0.26\textwidth]{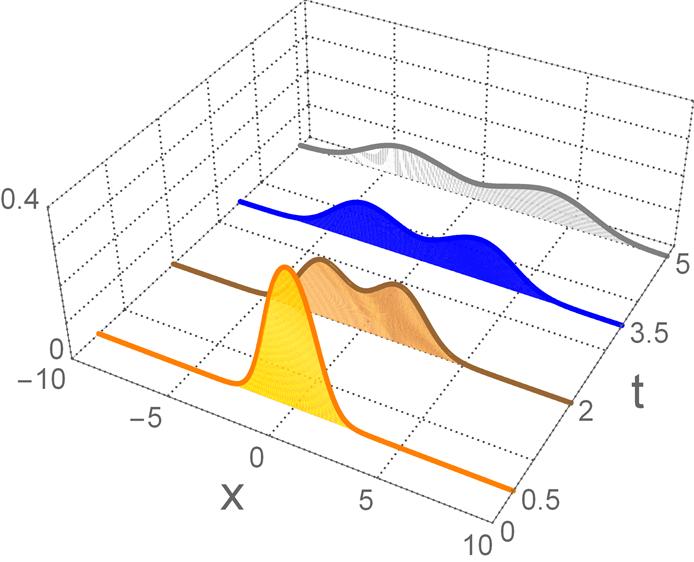} } 
\hskip.5ex
\subfigure[  $(1.9, 1.95 )$ ]{\includegraphics[width=0.26\textwidth]{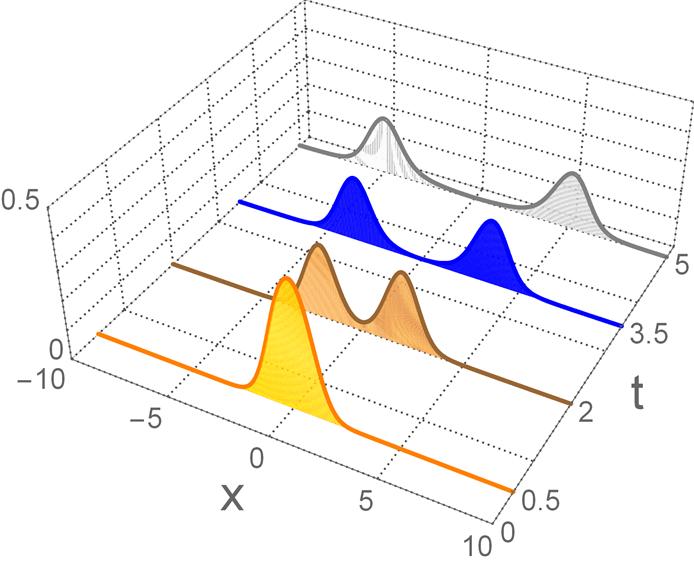} }

\subfigure[  $(1.1, 1.5 )$ ]{\includegraphics[width=0.26\textwidth]{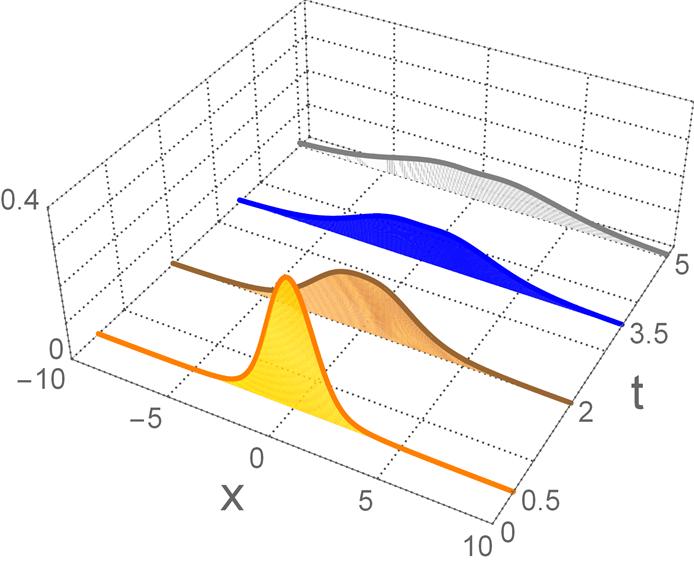} } 
\hskip.5ex
\subfigure[ $(1.5, 1.5 )$ ]{\includegraphics[width=0.26\textwidth]{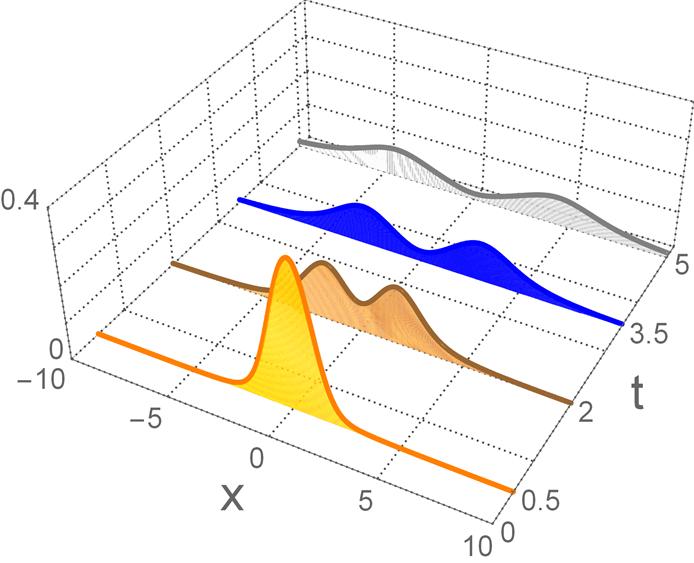} } 
\hskip.5ex
\subfigure[ $(1.9, 1.5 )$ ]{\includegraphics[width=0.26\textwidth]{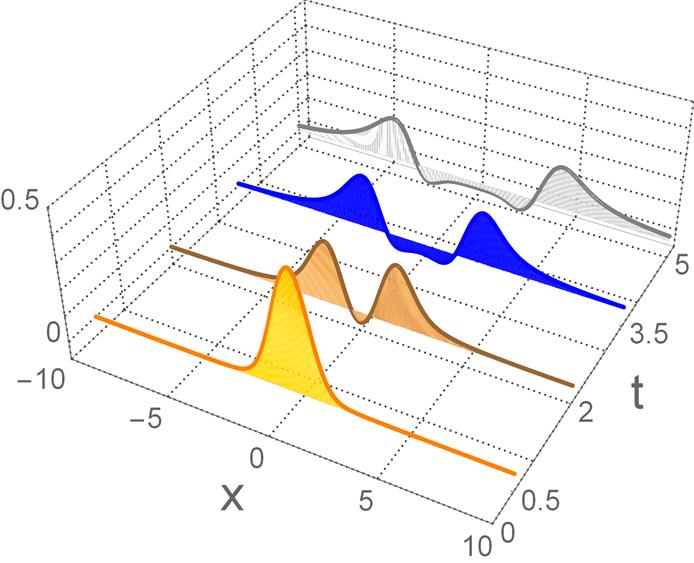} } 

\subfigure[  $(1.1, 1.05 )$ ]{\includegraphics[width=0.26\textwidth]{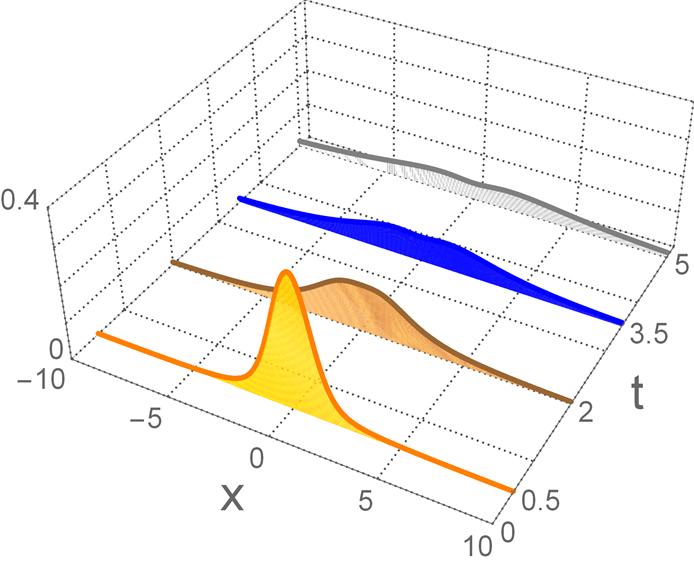} } 
\hskip.5ex
\subfigure[ $( 1.5, 1.05)$ ]{\includegraphics[width=0.26\textwidth]{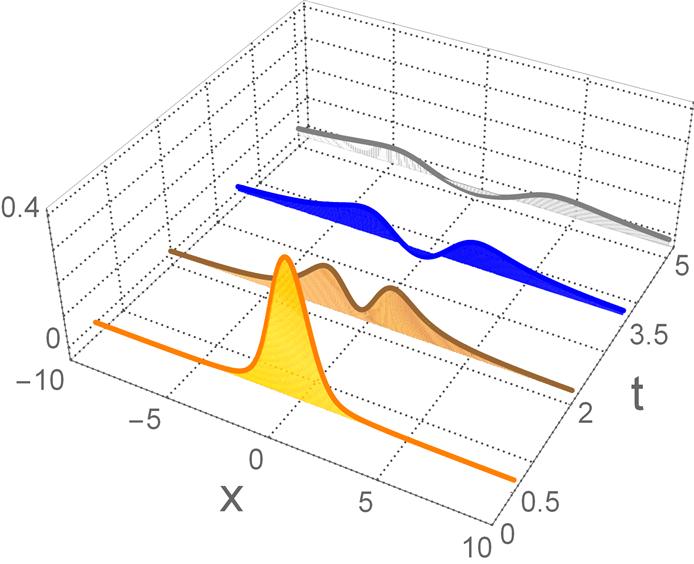} } 
\hskip.5ex
\subfigure[ $( 1.9, 1.05)$ ]{\includegraphics[width=0.26\textwidth]{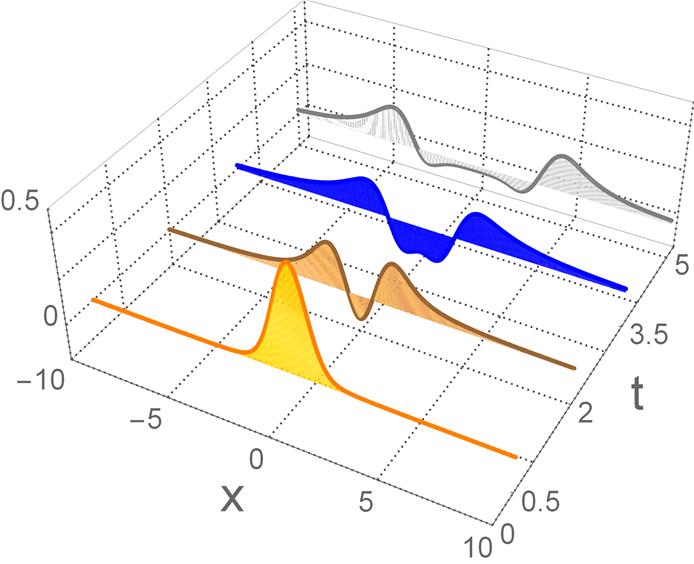} } 

\caption{\footnotesize Time-evolution of the functions $u_e(x,t)$ defined in (\ref{solg1})-(\ref{solg2}) for points $(\alpha, \beta)$ inside the grey squared area of Figure~\ref{plane}. The clockwise oriented sets of figures (a)--(c), (c)--(i), (i)--(g), and (g)--(a) follow the  segment-lines $I$--$IV$, respectively, and the set  (g), (e), (c) follows Segment-line $V$. In turn, Figures (a), (c), (i) and (g) refer to points $(\alpha, \beta)$ that are very close to the vertices $A$, $B$, $D$ and $C$, respectively. 
}
\label{gauss}
\end{figure}

\begin{figure}[htb]

\centering
\subfigure[ $(1.1, 1.95 )$ ]{\includegraphics[width=0.26\textwidth]{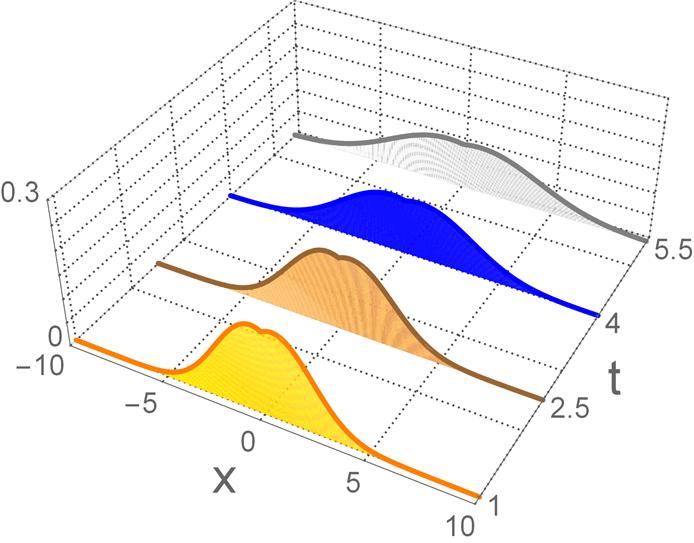} } 
\hskip.5ex
\subfigure[  $(1.5, 1.95 )$ ]{\includegraphics[width=0.26\textwidth]{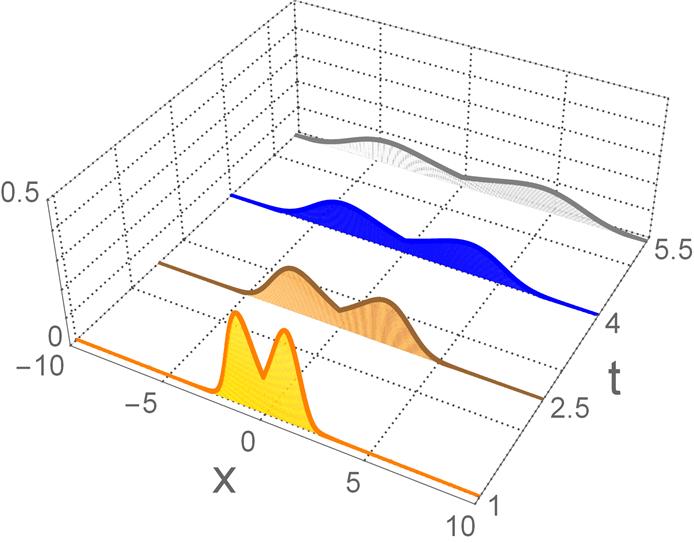} } 
\hskip.5ex
\subfigure[  $(1.9, 1.95 )$ ]{\includegraphics[width=0.26\textwidth]{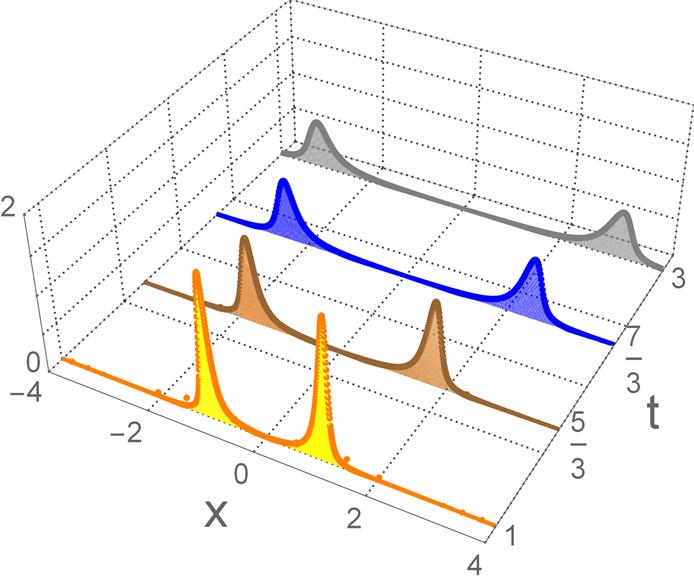} }

\subfigure[  $(1.1, 1.5 )$ ]{\includegraphics[width=0.26\textwidth]{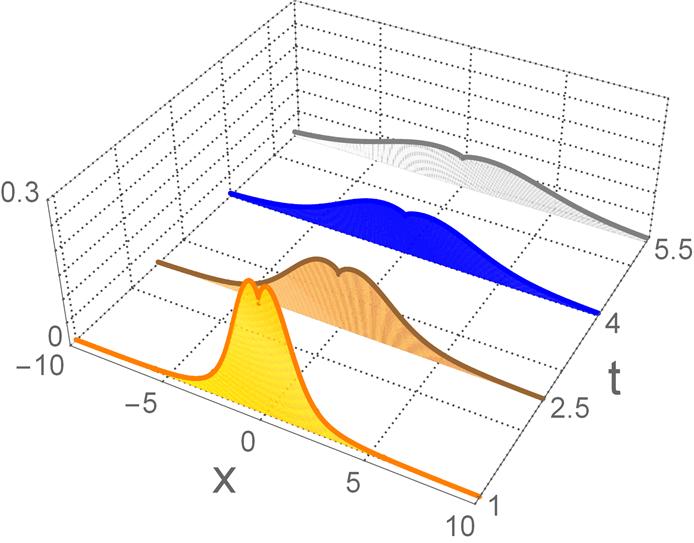} } 
\hskip.5ex
\subfigure[  $(1.1, 1.05 )$ ]{\includegraphics[width=0.26\textwidth]{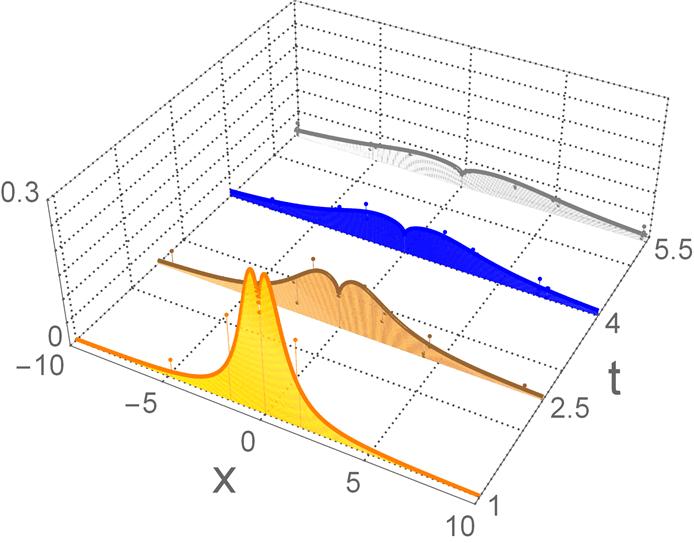} } 
\hskip.5ex
\subfigure[ $(1.5, 1.5 )$ ]{\includegraphics[width=0.26\textwidth]{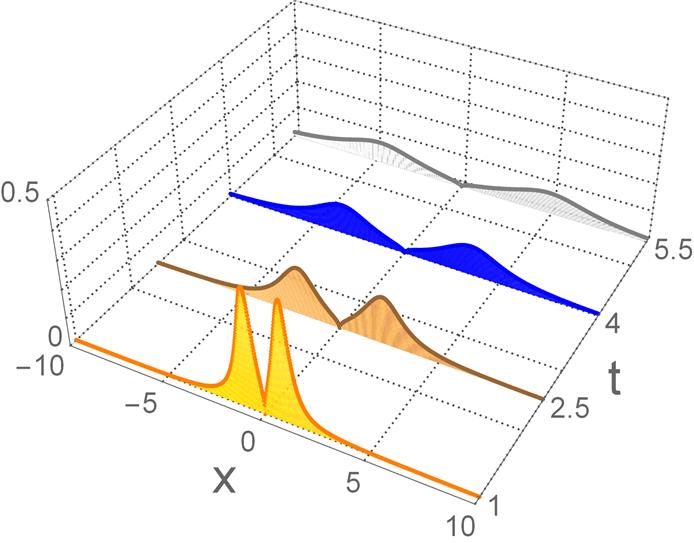} } 

\caption{\footnotesize Time-evolution of the functions $u(x,t)$ defined in (\ref{udelta}) for the indicated values of the pair $(\alpha,\beta)$ inside the circuit $BAD$ of Figure~\ref{plane}. The panel is clockwise oriented. Solutions in the upper row are very close to $u_I(x,t)$, defined in (\ref{u1}), and follow a path parallel to the segment-line $I$. Figures (c), (f) and (e) follow a path parallel to the segment-line $V$. The circuit is closed with Figures (e), (d) and (a), which follow a path parallel to the segment-line $IV$.
}
\label{fdelta1}
\end{figure}

As we have indicated in the previous section, using the Gaussian density (\ref{inig}) as initial condition is,  in many respects, more realistic than using the Dirac delta distribution (\ref{ini4}). The main point is that $\varphi_1^{(G)}(x)$ is finite at $x=0$ while $\varphi_1^{(\delta)}(x)$ diverges as $\vert x \vert \rightarrow 0$. However, the disadvantage of $\varphi_1^{(G)}(x)$ over $\varphi_1^{(\delta)}(x)$ is that the former, although centered at $x=0$, spreads along the entire $x$-axis while the latter is very localized at $x=0$ and equal to zero for any $x\neq 0$. Nevertheless, $\varphi_1^{(\delta)}(x)$ may be recovered as a Gaussian density with infinitesimally-narrow width. In the panel of Figure~\ref{gauss} we show the propagation of the initial disturbance $\varphi_1^{(G)}(x)$, according to the space-time fractional differential equation (\ref{frac1}), for different points $(a,b)$ in the grey squared area of Figure~\ref{plane}. The plots in Figure~\ref{gauss} are in correspondence with those shown in Figure~\ref{fdelta1}, where we have depicted the propagation of the Dirac delta-like initial disturbance $\varphi_1^{(\delta)}(x)$. That is, the behavior of the functions shown in Figure~\ref{fdelta1} may be interpreted as the form in which the functions of Figure~\ref{gauss} should behave at the limit $x_0 \rightarrow 0^+$.

The panel shown in Figure~\ref{gauss} includes the configuration for nine points $(\alpha, \beta)$ of the grey square of Figure~\ref{plane}. We would distinguish four different situations, with Fig.~\ref{gauss}(e) at the barycenter. In clockwise orientation, the behavior of the group formed by figures (a-c) and (e) exhibits a combination of wave propagation with diffusion. The former is stronger in (c), where $(\alpha, \beta)$ is closest to vertex $B$, and the latter is stronger in (a), where $(\alpha, \beta)$ is closest to vertex $A$. The clearest feature of the wave-propagation is the decoupling of the initial disturbance into two perturbations, the maxima of which evolve in time according to the respective characteristic integrals. As indicated above, these integrals are the constants $x \pm vt$ for vertex $B$. In our case, the characteristic integrals are not straight-lines (see below for details). In turn, diffusion is characterized by the fading of the initial disturbance as time goes pass (conservation principles imply that the disturbance must spread out with time in order to preserve the area under the initial curve $\varphi_1$). The behavior of the functions depicted in figures (b) and (e) shows, with clarity, a mixture of diffusion and wave propagation.

\begin{figure}[htb]
\centering 
\includegraphics[width=0.3\textwidth]{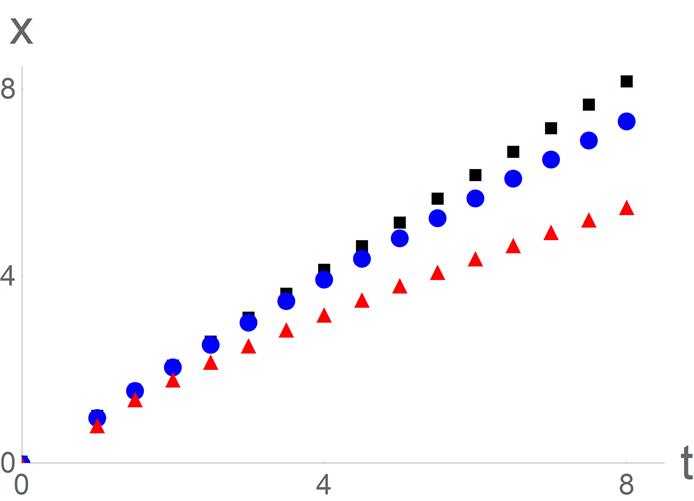} 

\caption{\footnotesize Propagation of the (right hand side) maxima of the solutions to the time-fractional differential equation (\ref{fracwe}) with initial conditions (\ref{inig}) and $\varphi_2=0$. We have taken $u_e(x,t)$ defined in (\ref{solg1})-(\ref{solg2}) with $\beta=2$, see Figure~\ref{gauss}.
The plots correspond to $\alpha=1.9$ (solid square, black), $\alpha=1.7$ (disk, blue) and $\alpha=1.5$ (solid triangle, red).
}
\label{recta1}
\end{figure}

A second group, formed by figures (c-i) and (e), shows the split of the initial disturbance into two perturbations that is characteristic of the wave-propagation. Additionally, the perturbations take negative values producing the presence of nodes as they propagate. The latter is markedly notorious in figure (i), where $(\alpha, \beta)$ is closest to vertex $C$. As we have shown, such behavior is a trait of the solutions of the complementary equation (\ref{Ceq}). Therefore, the behavior of functions included in this group is a mixture of wave- and complementary-propagation.

We have two additional groups, the behavior of which exhibits respectively a mixture of transport and complementary-propagation, and a mixture of diffusion and transport processes. The former group includes figures (i-g) and (e), and the latter is integrated by figures (g-a) and (e).

\begin{figure}[htb]
\centering 
\includegraphics[width=0.3\textwidth]{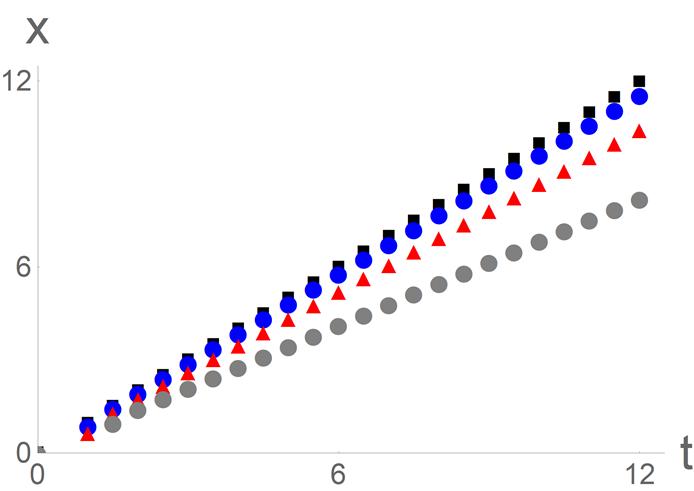} 

\caption{\footnotesize Propagation of the (right hand side) maxima of the functions $u_e(x,t)$ defined in (\ref{solg1})-(\ref{solg2}) for $\alpha=\beta$, see Figure~\ref{gauss}.
The plots correspond to $\alpha=1.9$ (solid square, black), $\alpha=1.7$ (disk, blue), $\alpha=1.5$ (solid triangle, red), and $\alpha=1.3$ (disk, grey).
}
\label{recta2}
\end{figure}

With respect to the characteristic integrals, in Figure~\ref{recta1} we show the distribution of the maxima $x_M(t)$ of $u_e(x,t)$ in the plane $t \times x$ for $\beta=2$ and three different values of $\alpha$. That is, Figure~\ref{recta1} refers to the propagation of the Gaussian density $\varphi_1^{(G)}(x)$ defined by the points $(\alpha, \beta)$ along the segment-line $I$. As we can see, for $\alpha =1.9$ the characteristic integral defines a path that is almost a straight-line, which might be expected since $(\alpha, \beta)= (1.9,2)$ is very close to vertex $B$, which corresponds to the wave equation (see the pretty explanation of the behavior of characteristic integrals for wave equation in \cite{Bor18}). On the other hand, the path defined by  $(\alpha, \beta)= (1.5,2)$ is far away from a straight-line. The results depicted in Figure~\ref{recta1} have been calculated numerically, and obey a rule (obtained by the best-fit technique) of the form
\[
x_M (t; \alpha, \beta=2) = \pm c_{\alpha} t^{\frac{\alpha}{2} + \theta_{\alpha} }, \quad 1 \leq \alpha \leq 2,
\]
where $c_{\alpha}$ and $\theta_{\alpha}$ are parameterized by $\alpha$. In contrast with other works like \cite{Fuj90}, we have found that $\theta_{\alpha} \neq 0$ and $c_{\alpha} \neq 1$ for $1 < \alpha <2$. Of course, $c_1=0$ and $c_2=1$, with $\theta_2=0$.

On the other hand, in Figure~\ref{recta2} we show the behavior of the maxima $x_M(t)$ for the solutions defined along the segment-line $V$. In this case, we have found the rule
\[
\widetilde x_M (t; \alpha = \beta) = \widetilde c_{\alpha} t^{1 + \tau_{\alpha} }, \quad 1 \leq \alpha \leq 2,
\]
with $\widetilde c_{\alpha}$ and $\tau_{\alpha}$ determined by $\alpha$. The value of $\widetilde x_M (t)$ is very close to that of $x_M(t)$  for $\alpha=1.9$, as expected. As in the previous case, unlike other works \cite{Fuj90}, in our approach the rule obeyed by the maxima is not as simple as $t^{\alpha/2}$.

To conclude our analysis, the panel of Figure~\ref{fdelta1} shows the behavior of (\ref{udelta}) for six different points $(\alpha, \beta)$. They correspond, in clockwise direction, to the functions depicted in Figures \ref{gauss}(a)--\ref{gauss}(c), \ref{gauss}(e), \ref{gauss}(g), and \ref{gauss}(d), respectively. Here, Figure \ref{fdelta1}(f) is the barycenter. Unlike the Gaussian case, in the group exhibiting propagation-diffusion behavior, Figures \ref{fdelta1}(a)--\ref{fdelta1}(c) and \ref{fdelta1}(f), the wavelike propagation is markedly differentiated at short times. Notice that the transport-diffusion processes are also ``accelerated'' in the sense that changes are presented in intervals of time that are shorter than those spent by the gaussian profiles. A similar phenomenon is observed for the other groups (not included in Figure~\ref{fdelta1}).

\section{Concluding remarks}
\label{conclu}

\begin{figure}[htb]
\centering 
\includegraphics[width=0.2\textwidth]{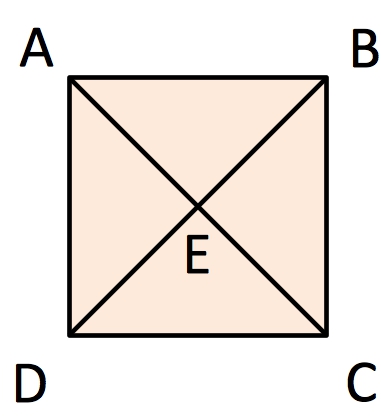} 

\caption{\footnotesize The square $1 \leq \alpha, \beta \leq 2$ of Figure~\ref{plane} with an additional segment-line $VI$ that connects the vertices $A$ and $C$. The addition of vertex $E$ splits the square into four different triangular areas that serve to classify the solutions of the space-time fractional differential equation (\ref{frac1}) according to their properties. Clockwise oriented, with vertex $E$ as barycenter, points $(\alpha, \beta)$ within the area delimited by the circuit $ijE$ define solutions with combined properties $i$ and $j$, where $i$ and $j$ stand for $A$ (heat diffusion), $B$ (wave propagation), $C$ (complementary propagation) and $D$ (transport process). See discussion in Section~\ref{results}.
}
\label{plane2}
\end{figure}

The discussion of Section~\ref{results} motivates the addition of a new segment-line  in Figure~\ref{plane}, named $VI$, connecting the vertices $A$ and $C$. Thus, $VI$ defines the simplest path to transit from the heat equation (\ref{he}) to the complementary equation (\ref{Ceq}). This can be written as the rule $\beta=3-\alpha$ for $1\leq \alpha \leq 2$. The presence of the new vertex $E$ splits the square into four different triangular areas, respectively bordered by the circuits $ABE$, $BCE$, $CDE$, and $DAE$, see Figure~\ref{plane2}. These regions of the $\alpha \times \beta$ plane are in correspondence with the groups mentioned in Section~\ref{results}, so they serve to classify the solutions of the space-time fractional differential equation (\ref{frac1}). Namely, solutions associated to points within the region $ABE$ will exhibit combined properties of wavelike propagation and diffusion. Those associated to $BCE$ will show a mixture of wavelike and complementary propagation, and so on.

We have shown that the $H$-function permits to express, in unified form, a family of solutions to the Cauchy problem defined by the space-time fractional differential equation (\ref{frac1}) and the initial conditions (\ref{ini1}). We have specialized to the case $\varphi_2=0$ for two different forms of $\varphi_1$. To be concrete, we have solved the problem for the Dirac delta pulse (\ref{ini4}) and for the Gaussian density (\ref{inig}) as well. We have shown that the solutions of the former can be obtained from those of the latter at the appropriate limit. In contrast with the Dirac delta distribution, which is recurrently used as initial condition in different fractional approaches, the Gaussian density is rarely included as initial condition in the literature on the matter. The material included in this work is addressed to fill such lack of information.

The well known solutions of three important equations studied in mathematical physics have been recovered as particular cases. We refer to the wave equation (\ref{we}), the heat equation (\ref{he}), and the transport equation (\ref{te}), which respectively correspond to vertices $B$, $A$, and $D$ of Figure~\ref{plane}, see also Figure~\ref{plane2}. In addition, we have completed the set of conic-type second-order partial differential equations
\[
F(x,t,u,u_x, u_t, u_{xx}, u_{tt})=0
\]
by including the differential equation that results from making $\alpha=2$ and $\beta=1$ in the space-time fractional differential equation (\ref{frac1}). That is, we have solved also the differential equation associated to vertex $C$, the explicit form of which is introduced in (\ref{Ceq}) and has been called {\em complementary equation} throughout the present work. In this form, besides including the fractional cases where $(\alpha, \beta)$ does not coincide with any vertex, our approach unifies in a single expression the solutions of four different problems that are usually treated in separate form in mathematical-physics.

Collateral results include the evaluation of different forms of the $H$-function. As far as we know, most of them have been unclassified up to now in the specialized literature. Particularly, the expressions of the $H$-functions in which the Dirac delta distribution is involved. We hope these results will be useful for researchers in the area.

The approach presented here may be extended to include nontrivial initial velocities $\varphi_2 \neq 0$. Specifically, the Dirac delta distribution $\varphi_2^{(\delta)}(x) = \gamma \delta(x - x_0)$ offers the possibility to calculate the Green functions for the problems above mentioned. Namely, it is well known that the calculation of the Green function for a given Cauchy problem is equivalent to the calculation of the solutions to the homogeneous equation with the boundary conditions $\varphi_1(x)=0$ and $\varphi_2^{(\delta)}(x)$ \cite{Duf15}. The latter works well for vertices $A$, $B$ and $D$, which produce conventional partial differential equations. The verification of such a property for vertex $C$ and any other point $(\alpha, \beta)$ in the squared area $1 \leq \alpha, \beta \leq 2$ is an open problem that we shall face elsewhere.

To conclude this work we would like to emphasize that the substitution of conventional derivatives by their fractional versions in a given dynamical law produces the emergence of interactions that are not apparent  (and cannot be noticed) in conventional models. The framework developed here look for connections between different laws that are already known. But fractional calculus offers a more wide range of possibilities. For example, in the case of the harmonic oscillator, the fractional time-derivative adds a sort of frictional forces that are not justified in the conventional Newtonian model since no environmental interactions are assumed a priori \cite{Oli17}. Therefore, the fractional version of the oscillator presupposes either that the system suffers a kind of self-interaction or that it is embedded into a medium with memory. Both assumptions predict phenomena that may require a new dynamical law for their explanation. A similar situation occurs for the fractional quantum oscillator \cite{Oli16}, for which there exist some immediate applications in quantum optics \cite{Hua17}. Work in this direction is in progress.

\appendix
\section{Useful definitions and expressions}
\label{ApA}

\renewcommand{\thesection}{A-\arabic{section}}
\setcounter{section}{0}  

\renewcommand{\theequation}{A-\arabic{equation}}
\setcounter{equation}{0}  

$\bullet$ The Mellin transform \cite{Ber00} of a function $f(x)$ is defined as
\be
f_{\cal M}(s) ={\cal M} [f(x)]  = \int_0^{\infty} f(x) x^{s-1} dx.
\label{a1}
\ee
Consistently, the Mellin-Barnes integral (inverse Mellin transformation) is written as follows
\be
{\cal M}^{-1} [ f_{\cal M}(s) ] = \frac{1}{2\pi i} \int_{c-i\infty}^{c+\infty} f_{\cal M}(s) x^{-s} ds = f(x).
\ee
In particular, for $f(x) = \delta (x-x_0)$ one has
\be
{\cal M} [\delta(x -x_0)] = x_0^{s-1}, \qquad x_0 \delta(x-x_0) = \frac{1}{2\pi i}  \int_{c-i\infty}^{c+\infty}  \left( \frac{x}{x_0}
\right)^{-s} ds, \quad x_0 >0.
\label{a3}
\ee

\noindent
$\bullet$ The Fox function $H_{p,q}^{m,n}[x \vert -]$, $H$-function for short, is defined by the Mellin-Barnes integral 
\be
H_{p,q}^{m,n} \left[ z \left\rvert 
\begin{array}{cc} 
(a_{1},A_{1}), \ldots, (a_{p},A_{p}) \\[1ex] 
(b_{1},B_{1}), \ldots, (b_{q},B_{q})  
\end{array} 
\right. \right] = \frac{1}{2\pi i} \int_L \Lambda(s) z^{-s} ds,
\label{H}
\ee
where $m, n, p$, and $q$ are nonnegative integers, $a_{i},b_{j}\in\mathbb{C}$, $A_{i},B_{j}\in(0, \infty)$, and 
\be
\Lambda(s) = 
\frac{ \prod_{j=1}^{m} \Gamma(b_{j} +B_{j} s)  \prod_{i=1}^{n} \Gamma(1-a_{i}-A_{i} s)  }{ \prod_{j=m+1}^{q} \Gamma(1-b_{j}- B_{j} s) \prod_{i=n+1}^{p} \Gamma(a_{i}+ A_{i} s ) }, \quad 1 \leq m \leq q, \quad 0 \leq n \leq p.
\label{H2}
\ee
The contour $L$ in (\ref{H}) separates the poles of $\Gamma (b_j -B_j s)$, $j=1,\ldots,m$, from the poles of $\Gamma (1-a_i +A_i s)$, $i=1, \ldots, n$. Detailed information can be found in \cite{Mat10}.

\noindent
$\bullet$ Consider the quantity
\be
\Delta=\sum_{j=1}^{q}\beta_{j}-\sum_{i=1}^{p}\alpha_{i}.
\ee
The following theorems are reproduced from Ref.~\cite{Kil04}.

\textbf{Theorem 1.3.} \cite{Kil04}: Provided $\Delta > 0$ and $z\neq0$, the H-function can be expanded as the series
\be
H_{p,q}^{m,n}(z) = \sum_{j=1}^{m} \sum_{\ell = 0}^{\infty} h_{j \ell}^*  z^{(b_j + \ell )/\beta_j } ,
\ee
where
\be
h_{j \ell}^* = \frac{(-1)^{\ell} }{\ell ! \beta_j } \frac{ \prod_{i=1,i\neq j}^{m} \Gamma \left(b_{i}-[b_{j} + \ell ]\frac{\beta_i }{\beta_j } \right) \prod_{i=1}^{n} \Gamma \left( 1-a_{i} + [b_{j} + \ell ] \frac{\alpha_i }{\beta_j} \right ) }{ \prod_{ i = n+1}^{p} \Gamma \left( a_{i} - [b_{j} + \ell ] \frac{ \alpha_{i} }{ \beta_{j} }\right) \prod_{i=m+1}^{q} \Gamma \left(1-b_{i}+[b_{j}+ \ell] \frac{\beta_{i}}{\beta_{j}}\right)},
\ee
and
\bea
\beta_{j}(b_{i}+k)\neq\beta_{i}(b_{j}+ \ell), \quad i \neq j; \quad i,j =1,...,m; \quad k,\ell = 0,1,2, \ldots,
\label{t1} \\[2ex]
\alpha_{i}(b_{j}+ \ell )\neq\beta_{j}(a_{i}-k-1), \quad i=1,...,n;  \quad j = 1,...,m; \quad k, \ell = 0,1,2, \ldots 
\label{t1}
\eea

\textbf{Theorem 1.4.} \cite{Kil04}: Provided $\Delta < 0$ and $z\neq0$, the H-function can be expanded as the series

\be
H_{p,q}^{m,n}(z)=\sum_{j=1}^{n}\sum_{k=0}^{\infty}h_{ik}z^{(a_{j}-1-k)/\alpha_{i}}, 
\ee
where
\be 
h_{ik}=\frac{(-1)^{k}}{k!\alpha_{i}}\frac{\prod_{j=1}^{m}\Gamma\left(b_{j}+[1-a_{i}+k]\frac{\beta_{j}}{\alpha_{i}}\right)\prod_{j=1,j\neq i}^{n}\Gamma\left(1-a_{j}-[1-a_{i}+k]\frac{\alpha_{j}}{\alpha_{i}}\right)}{\prod_{j=n+1}^{p}\Gamma\left(a_{j}+[1-a_{i}+k]\frac{\alpha_{j}}{\alpha_{i}}\right)\prod_{j=m+1}^{q}\Gamma\left(1-b_{j}-[1-a_{i}+k]\frac{\beta_{j}}{\alpha_{i}}\right)}
\ee
and, besides (\ref{t1}), the following constrain is also true
\be
\alpha_{j}(1-a_{i}+k) \neq \alpha_{i}(1-a_{j}+l), \quad i\neq j;  \quad i,j=1,...,n; \quad k,l=0,1,2, \ldots
\ee

\noindent
$\bullet$ The following expression connects the $H$-function with the so-called Wright function
\bea
H_{1,1}^{1,0} \left[ z  \left\rvert 
\begin{array}{cc} 
(a,A) \\
(0,1)
\end{array} 
\right.
\right] &=&  \frac{1}{2 \pi i} \int_L \frac{ \Gamma( z)}{ \Gamma(a+A s) } z^{-s} ds\\[1ex]
&=& {}_0 \Psi_1 \left[ -z \left\vert 
\begin{array}{c}
-\\
(a,-A)
\end{array}
\right. \right] \equiv \phi(-A,a; -z).
\eea
The $H$-function and the generalized Wright function ${}_p \Psi_q [z]$ are related as follows
\be
H_{p,q+1}^{1,p} \left[ z \left\rvert 
\begin{array}{cc} 
(1-a_{1},A_{1}), \ldots, (1-a_{p},A_{p}) \\[1ex] 
(0,1), (1- b_{1},B_{1}), \ldots, (1- b_{q},B_{q})  
\end{array} 
\right. \right] = {}_p \Psi_q \left[ - z \left\rvert 
\begin{array}{cc} 
(a_p ,A_p ) \\[1ex] 
(b_q,B_q) 
\end{array} 
\right. \right].
\label{HPsi}
\ee

\noindent
$\bullet$ The duplication formula of the Gamma function is given by
\be
\Gamma(z) \Gamma (z + \tfrac{1}{2} )=2^{1-2z} \sqrt{\pi} \Gamma (2 z).
\label{dupli}
\ee

\noindent
$\bullet$ The Cahen-Mellin integral \cite{Har16} is defined as 
\be
\frac{1}{2\pi i}\int_{L}\Gamma(s)y^{-s}ds=e^{y}.
\label{CMf}
\ee

\noindent
$\bullet$ The following expression is useful to derive Eq.~(\ref{serie2}) in Section~\ref{discussion}:
\[
\Gamma ( \tfrac12 - m) = \frac{ (-4)^m m! \sqrt{\pi}}{ (2m)! }.
\]

\appendix
\setcounter{section}{1}
\section{Derivation of Eq.~(\ref{udes}) }
\label{ApB}

\renewcommand{\theequation}{B-\arabic{equation}}
\setcounter{equation}{0}  

The integrand in Eq.~(\ref{lap2}) is an even function of $k$, so we can write
\be 
U(x,s) = \frac{\mu \lambda}{ \pi s} \int_{0}^{\infty} \frac{ \cos( \lambda y x)}{1+y^{\beta}} dy,
\label{b1}
\ee
with
\be
y = \frac{k}{\lambda}, \quad \lambda = s^{\alpha/\beta} v_{\alpha,\beta}^{-2/\beta}.
\label{b2} 
\ee
The calculation of (\ref{b1}) is facilitated by the Mellin transform (\ref{a1}) of $U(x,s)$. Namely
\bea
{\cal M} [U(x,s)]  & = & \int_0^{\infty} U(x,s) x^{z-1} dx 
\label{b31}\\[2ex]
& = & \left( \frac{\mu \lambda^{1-z} }{s} \right) \frac{ \Gamma(z)}{ \Gamma \left( \frac{1-z}{2} \right) \Gamma \left( \frac{1+z}{2} \right)  } \displaystyle\int_0^{\infty} \frac{y^{-z} }{ 1 + y^{\beta} } dy,
\label{b3}
\eea
where we have used
\be
{\cal M}[ \cos(wx)]= \left[ \frac{ \pi \Gamma(z) }{ \Gamma \left(\frac{1-z}{2} \right)\Gamma \left( \frac{1+z}{2} \right)} \right] w^{-z}.
\label{bcos}
\ee
The integral in (\ref{b3}) may be determined by Eq.~3.241.2 of \cite{Gra96} to arrive at the moment problem
\be
 \int_0^{\infty} U(x,s) x^{z-1} dx = \mu \left[ \frac{ \Gamma(z) \Gamma \left( 1 + \frac{z-1}{\beta} \right) \Gamma \left( \frac{1-z}{\beta}
 \right) }{ \Gamma \left(\frac{1-z}{2} \right)\Gamma \left( \frac{1+z}{2} \right)} \right]  \left( \frac{\lambda^{1-z} }{s} \right),
\label{b4}
\ee
which is convergent for $0 < \frac{1- z}{\beta} <1$. According to Eqs.~(\ref{H}) and (\ref{H2}) of Appendix~\ref{ApA}, the above expression corresponds to the Mellin-Barnes integral representation of an $H$-function. To be concrete
\[
U(x,s)= \left(  \frac{ \mu \, s^{\alpha/\beta-1} }{ \beta \, v_{\alpha,\beta}^{2/\beta}} \right) 
H_{2,3}^{2,1} \left[ \left( \frac{s^{\alpha/\beta}}{v_{\alpha,\beta}^{2/\beta}} \right) \vert x \vert \,\biggr\rvert 
\begin{array}{cc} 
\left( \frac{\beta -1}{\beta},\frac{1}{\beta} \right), \left( \frac{1}{2},\frac{1}{2} \right) \\[1.5ex]
(0,1), \left( \frac{\beta -1}{\beta},\frac{1}{\beta} \right), \left( \frac{1}{2},\frac{1}{2} \right) 
\end{array}
\right], 
\]
which has been included as Eq.~(\ref{udes}) in section~\ref{iniciales1}.

\appendix
\setcounter{section}{2}
\section{Derivation of Eq.~(\ref{udelta})}
\label{ApC}

\renewcommand{\theequation}{C-\arabic{equation}}
\setcounter{equation}{0}  

The inverse Laplace transform of $U(x,s)$ is calculated at the elementary level by noticing that the $s$-variable, which is to be transformed, is encoded in the term $\lambda^{1-z}/s$ appearing as a factor on the right hand side of (\ref{b4}). From (\ref{b2}) we have
\[
\frac{\lambda^{1-z}}{s} =  v_{\alpha, \beta}^{-2 \left( \frac{1-z}{\beta} \right) } \, s^{\alpha \left( \frac{1-z}{\beta} \right) -1}.
\]
Therefore
\[
\mathfrak{L}^{-1} \left[ s^{\alpha \left( \frac{1-z}{\beta} \right) -1} \right] = \frac{t^{-\frac{\alpha(1-z)}{\beta}}}{\Gamma\left(1-\frac{\alpha(1-z)}{\beta}\right)}, \qquad \alpha ( \tfrac{1-z}{\beta} ) <1,
\]
permits to express the solution $u(x,t)$ in the Mellin-Barnes integral representation
\be 
u(x,t)= \left( \frac{\mu}{\beta v_{\alpha,\beta}^{2/\beta} \, t^{\frac{\alpha}{\beta}}} \right) 
\frac{1}{2\pi i} \displaystyle\int_{L} \frac{ \Gamma (z ) \Gamma \left( 1+  \frac{ z-1 }{\beta} \right)\Gamma \left( \frac{1-z}{\beta} \right) }
{ \Gamma \left( \frac{1-z}{2} \right) \Gamma \left( \frac{1+z}{2} \right) \Gamma \left( 1 -\frac{  \alpha(1-z)}{\beta} \right) } 
\left( \frac{x}{ v_{\alpha,\beta}^{2/\beta} \, t^{\frac{\alpha}{\beta}} } \right)^{-z} dz. 
\label{c1}
\ee
From Eqs.~(\ref{H}) and (\ref{H2}) of Appendix~\ref{ApA} we finally obtain
\[ 
u(x,t)= \left( \frac{\mu}{\beta \, t^{\frac{\alpha}{\beta} } \, v_{\alpha,\beta}^{2/\beta} } \right) H_{3,3}^{2,1} \left[ \frac{ \vert x \vert }{t^{\frac{\alpha}{\beta} } \, v_{\alpha,\beta}^{2/\beta} } \,  \biggr\rvert 
\begin{array}{cc} 
\left( \frac{\beta -1}{\beta}, \frac{1}{\beta} \right),  \left( \frac{1}{2}, \frac{1}{2} \right), \left( \frac{ \beta - \alpha}{\beta}, \frac{\alpha}{\beta} \right) \\[1.5ex]
(0,1), \left( \frac{\beta - 1}{\beta},\frac{1}{\beta} \right), \left( \frac{1}{2}, \frac{1}{2} \right) 
\end{array} 
\right],
\]
which corresponds to Eq.~(\ref{udelta}) of section~\ref{iniciales1}.

\appendix
\setcounter{section}{3}
\section{Derivation of Eqs.~(\ref{solg1})-(\ref{solg2})}
\label{ApD}

\renewcommand{\theequation}{D-\arabic{equation}}
\setcounter{equation}{0}  

For $\varphi_1(x)$ given in (\ref{inig}) and $\varphi_2(x) =0$, the integral (\ref{lap1}) is simplified as follows
\be
U(x,s) = \frac{\mu \lambda}{ \pi s} \int_{0}^{\infty} \frac{ \cos( \lambda y x)}{1+y^{\beta}} \exp \left(  -\frac{x_0^2 \lambda^2 y^2}{4} \right) dy,
\label{d1}
\ee
where we have used Eq.~(\ref{b2}) of Appendix~\ref{ApB}. Now, assisted by Eq.~(\ref{bcos}) of Appendix~\ref{ApB}, we calculate the Mellin transform of (\ref{d1}) to get
\be
{\cal M} \left[ U(x,s) \right] = \frac{ \mu \lambda^{1-z} \Gamma(z) }{s \Gamma \left( \frac12 -\frac12 z \right)\Gamma \left( \frac12 + \frac12 z \right) } \int_{0}^{\infty} \frac{ e^{-x_0^2 \lambda^2 y^2/4 } y^{-z} }{1+y^{\beta} } dy.
\label{d2}
\ee
After the power series expansion of the Gaussian factor in the integrand of (\ref{d2}), we use Eq.~3.241.2 of \cite{Gra96} to arrive at the moment problem
\be
{\cal M} [U(x,s)] =\frac{\mu\lambda^{1-z}}{ \beta s} \sum_{k=0}^{\infty} 
\frac{ \Gamma(z)\Gamma\left(1-\frac{1}{\beta}-\frac{2k}{\beta}+\frac{z}{\beta}\right)\Gamma\left(\frac{1}{\beta}+\frac{2k}{\beta}-\frac{z}{\beta}\right) } { \Gamma\left(\frac{1}{2}-\frac{1}{2}z\right)\Gamma\left(\frac{1}{2}+\frac{1}{2}z\right) k! } 
\left(-\frac{x_{0}^{2}\lambda^{2}}{4}\right)^k,
\ee
which is convergent for $0 < 1 +2k -z < \beta$. Calculating the inverse Laplace transform of the Mellin-Barnes integral of the above equation, with
\[
\mathfrak{L}^{-1} \left[ \frac{1}{s^{ 1-\frac{\alpha}{\beta}-\frac{2k\alpha}{\beta}+\frac{\alpha z}{\beta}  } } \right] =\frac{t^{-\frac{\alpha}{\beta}-\frac{2k\alpha}{\beta}+\frac{\alpha z}{\beta}}}{\Gamma(1-\frac{\alpha}{\beta}-\frac{2k\alpha}{\beta}+\frac{\alpha z}{\beta})}, 
\qquad 
1-\tfrac{\alpha}{\beta}- \tfrac{2k\alpha}{\beta} + \tfrac{\alpha z}{\beta} >0,
\]
we finally have
\[
u_e(x,t)= \frac{\mu}{ \beta t^{\frac{\alpha}{\beta} } \, v_{\alpha,\beta}^{2/\beta} } \sum_{k=0}^{\infty} \frac{(-1)^k }{k!} \left( \frac{x_0^2}{ 4 t^{\frac{2\alpha}{\beta} } \, v_{\alpha,\beta}^{4/\beta} } \right)^k  \Theta_k(z;\alpha,\beta),
\]
with
\[
\Theta_k(z;\alpha,\beta) = H_{3,3}^{2,1} \left[ \left. \frac{ \vert x \vert }{ 
t^{\frac{\alpha}{\beta} } \, v_{\alpha,\beta}^{2/\beta}  } \right\rvert 
\begin{array}{cc} 
\left( \frac{\beta -(1 + 2k)}{\beta}, \frac{1}{\beta} \right),
(\frac12, \frac12),
\left( \frac{\beta-\alpha (1+2k) }{\beta}, \frac{\alpha}{\beta} \right)\\[1.5ex]
(0,1), \left( \frac{\beta -(1+2k) }{\beta}, \frac{1}{\beta} \right), 
(\frac12, \frac12) 
\end{array}
\right].
\]
The above result is included as Eqs.~(\ref{solg1})-(\ref{solg2}) of Section~\ref{iniciales2}.

\section*{Acknowledgment}

The authors are grateful to Professor L.M.~Nieto for useful comments. The financial support from Spanish MINECO (Project MTM2014-57129-C2-1-P), Junta de Castilla y Le\'on (VA057U16), and CONACyT is acknowledged. F. Olivar-Romero gratefully acknowledges the funding received through the CONACyT Scholarship 397335, and thanks the hospitality provided by the Department of Theoretical Physics of the University of Valladolid, Spain, where part of this work has been done.


\end{document}